%% file: HIG-18-011_temp.tex
\pdfoutput=1

\documentclass[11pt,twoside,a4paper,cmspaper,final,collab]{cms-tdr}
\def\svnVersion{a3e5a60-D}\def\svnDate{2019/06/17}\def\cmsCernNoTag{CERN-EP-2018-309}\def\cmsCernDate{\today}\def\cmsMessage{"Published in Physics Letters B as \href{http://dx.doi.org/10.1016/j.physletb.2019.06.021}{\doi{10.1016/j.physletb.2019.06.021}.}"}

\begin{document}\cmsNoteHeader{HIG-18-011}

\hyphenation{had-ron-i-za-tion}
\hyphenation{cal-or-i-me-ter}
\hyphenation{de-vices}
\RCS$HeadURL$
\RCS$Id$

\newcommand{\il}{\ensuremath{35.9\fbinv}\xspace}
\newcommand{\PFrelIso}{\ensuremath{I_{\mathrm{rel}}}\xspace}
\newcommand{\pfChargedHadronIso}{\ensuremath{I^{\text{ch.\,h}}}\xspace}
\newcommand{\pfPhotonIso}{\ensuremath{I^{\cPgg}}\xspace}
\newcommand{\pfNeutralHadronIso}{\ensuremath{I^{\text{n.\,h}}}\xspace}
\newcommand{\pfPU}{\ensuremath{I^{\text{PU\,ch.\,h}}\xspace}}
\newcommand{\sigmaGGH}{\ensuremath{\sigma_{{\cPg\cPg}\to\Ph}}\xspace}
\newcommand{\sigmaVBF}{\ensuremath{\sigma_{{\cPq\cPq}\to\Ph}}\xspace}
\newcommand{\mH}{\ensuremath{125\,}\xspace}
\newcommand{\mfour}{\ensuremath{m_{\Pgm\Pgm\cPqb\cPqb}}\xspace}
\newcommand{\ma}{\ensuremath{m_{\Pa_1}}\xspace}
\newcommand{\ab}{\ensuremath{{\Pa_1}}\xspace}
\newcommand{\mmumu}{\ensuremath{m_{\Pgm\Pgm}}\xspace}
\newcommand{\mbb}{\ensuremath{m_{\cPqb\cPqb}}\xspace}
\newcommand{\myxsec}{\ensuremath{\sigma_{\Ph}{\mathcal{B}({\Ph}\to{\ab\ab}\to\MM{\bbbar})}}\xspace}
\newcommand{\mybr}{\ensuremath{\mathcal{B}({\Ph}\to{\ab\ab}\to\MM{\bbbar})}\xspace}
\newcommand{\lbound}{\ensuremath{20}\xspace}
\newcommand{\hbound}{\ensuremath{62.5}\xspace}
\newcommand{\chit}{\ensuremath{\chi^2}\xspace}
\newcommand{\chib}{\ensuremath{\chi_{\cPqb\cPqb}}\xspace}
\newcommand{\chih}{\ensuremath{\chi_{\Ph}}\xspace}
\newcommand{\mmbb}{\ensuremath{\MM\bbbar}\xspace}
\newcommand{\mmtt}{\ensuremath{\MM\TT}\xspace}
\newcommand{\bbtt}{\ensuremath{\bbbar\TT}\xspace}
\newcommand{\mandelt}{\ensuremath{t}\xspace}
\newcommand{\mandels}{\ensuremath{s}\xspace}
\newcommand{\minTotObs}{\ensuremath{5}\xspace}
\newcommand{\maxTotObs}{\ensuremath{33}\xspace}
\newcommand{\minTotExp}{\ensuremath{7}\xspace}
\newcommand{\maxTotExp}{\ensuremath{14}\xspace}
\newcommand{\minBrObsNum}{\ensuremath{1}\xspace}
\newcommand{\maxBrObsNum}{\ensuremath{7}\xspace}
\newcommand{\minBrExpNum}{\ensuremath{1}\xspace}
\newcommand{\maxBrExpNum}{\ensuremath{3}\xspace}
\newcommand{\minBrObs}{\ensuremath{\minBrObsNum\times10^{-4}}\xspace}
\newcommand{\maxBrObs}{\ensuremath{\maxBrObsNum\times10^{-4}}\xspace}
\newcommand{\minBrExp}{\ensuremath{\minBrExpNum\times10^{-4}}\xspace}
\newcommand{\maxBrExp}{\ensuremath{\maxBrExp\times10^{-4}}\xspace}
\newcommand{\mur}{\ensuremath{\mu_\mathrm{r}}\xspace}
\newcommand{\muf}{\ensuremath{\mu_\mathrm{f}}\xspace}
\providecommand{\CL}{CL\xspace}
\newlength\cmsFigWidth
\ifthenelse{\boolean{cms@external}}{\setlength\cmsFigWidth{0.49\textwidth}}{\setlength\cmsFigWidth{0.65\textwidth}}
\ifthenelse{\boolean{cms@external}}{\providecommand{\cmsLeft}{upper\xspace}}{\providecommand{\cmsLeft}{left\xspace}}
\ifthenelse{\boolean{cms@external}}{\providecommand{\cmsRight}{lower\xspace}}{\providecommand{\cmsRight}{right\xspace}}

\cmsNoteHeader{HIG-18-011}

\title{
Search for an exotic decay of the Higgs boson to a pair of light pseudoscalars in the final state with two muons and two \cPqb\ quarks in $\Pp\Pp$ collisions at 13\TeV
}

\date{\today}

\abstract{
  A search for exotic decays of the Higgs boson to a pair of light pseudoscalar particles $\ab$ is performed under the hypothesis that one of the pseudoscalars decays to a pair of opposite sign muons and the other decays to $\bbbar$. Such signatures are predicted in a number of extensions of the standard model (SM), including next-to-minimal supersymmetry and two-Higgs-doublet models with an additional scalar singlet. The results are based on a data set of proton-proton collisions corresponding to an integrated luminosity of \il, accumulated with the CMS experiment at the CERN LHC in 2016 at a centre-of-mass energy of 13\TeV. No statistically significant excess is observed with respect to the SM backgrounds in the search region for pseudoscalar masses from 20\GeV to half of the Higgs boson mass. Upper limits at 95\% confidence level are set on the product of the production cross section and branching fraction, $\sigma_{\Ph}\mathcal{B}(\Ph\to \ab\ab\to\MM\bbbar)$, ranging from \minTotObs to \maxTotObs\unit{fb}, depending on the pseudoscalar mass. Corresponding limits on the branching fraction, assuming the SM prediction for $\sigma_{\Ph}$, are $(\minBrObsNum$--$\maxBrObsNum)\times 10^{-4}$.}

\hypersetup{
pdfauthor={CMS Collaboration},
pdftitle={Search for an exotic decay of the Higgs boson to a pair of light pseudoscalars in the final state with two muons and two b quarks in pp collisions at 13 TeV},
pdfsubject={CMS},
pdfkeywords={CMS, physics, BSM Higgs physics}}

\maketitle

\section{Introduction}
\label{sect:intro}
The discovery of the particle now identified as the Higgs boson by the ATLAS and CMS experiments~\cite{Aad20121,Chatrchyan201230,Chatrchyan:2013lba} at the CERN LHC~\cite{JINSTLHC} has opened a new era in the history of particle physics. So far, precise measurements of the Higgs boson spin, parity, width, and couplings in production and decay have been consistent with the expectations for the standard model (SM) Higgs boson~\cite{Aaboud:2017vzb,Sirunyan:2017exp,ATLASCMSHIGGSCOMBRUNI,PhysRevLett.110.081803}. However, the possibility of exotic Higgs boson decays to new lighter bosons is not excluded, and is proposed in various theories beyond the SM (BSM)~\cite{PhysRevD.90.075004}. The LHC combination of the SM Higgs boson measurements at 7 and 8\TeV allows Higgs boson decays to BSM states with a rate of up to $34\%$~\cite{ATLASCMSHIGGSCOMBRUNI} at 95\% confidence level (\CL). The LHC data at 13\TeV have been used to place an upper limit of about 40\% for the Higgs boson branching fraction ($\mathcal{B}$) to BSM particles at 95\% \CL~\cite{CMS-PAS-HIG-17-031}.

Several searches for exotic decays of the Higgs boson have been performed at the LHC, using the data at 8\TeV~\cite{Aad:2015oqa,Khachatryan:2015wka,Khachatryan:2015nba,Khachatryan:2017mnf} and 13\TeV~\cite{Aaboud:2016oyb,Aaboud:2018fvk,Aaboud:2018gmx,Aaboud:2018iil,Aaboud:2018esj,CMS-PAS-HIG-17-024,CMS-PAS-HIG-17-029}. Such decays occur in the context of the next-to-minimal supersymmetric standard model, NMSSM, and other extensions to two-Higgs-doublet models (2HDM) where the existence of a scalar singlet is hypothesised (2HDM+S)~\cite{PhysRevD.90.075004,DINE1981199,Kim:1983dt,Ellwanger:2009dp}. The 2HDM, and hence 2HDM+S, are categorised into four types depending on the interaction of SM fermions with the Higgs doublet structure~\cite{Khachatryan:2017mnf}. All SM particles couple to the first Higgs doublet, $\Phi_1$, in type I models. In type II models, which include the NMSSM, up-type quarks couple to $\Phi_1$ while leptons and down-type quarks couple to the second Higgs doublet, $\Phi_2$. Quarks couple to $\Phi_1$ and leptons couple to $\Phi_2$ in type III models. In type IV models, leptons and up-type quarks couple to $\Phi_1$, while down-type quarks couple to $\Phi_2$. After electroweak symmetry breaking, the 2HDM predicts a pair of charged Higgs bosons \Hpm, a neutral pseudoscalar \PSA, and two neutral scalar mass eigenstates, \PH and \Ph.
In the decoupling limit the lighter scalar eigenstate, $\Ph$, is the observed boson with $m_{\Ph}\approx\mH\GeV$. In 2HDM+S models, a complex scalar singlet $S_{\mathrm{R}}+iS_{\mathrm{I}}$ that has no direct Yukawa couplings is introduced. Hence, it is expected to decay to SM fermions by virtue of mixing with the Higgs sector. This mixing is small enough to preserve the SM-like nature of the $\Ph$ boson.

In this Letter we consider the Higgs boson decay to a pair of \ab particles where \ab is a pseudoscalar mass eigenstate mostly composed of $S_{\mathrm{I}}$. We perform a search for the decay chain ${\Ph}\to\ab\ab\to\mmbb$. The gluon gluon fusion (\cPg\cPg F) and the vector boson fusion (VBF) production mechanisms are considered, with production cross sections of $48.58\pm 2.45\unit{pb}$ (at next-to-next-to-next-to-leading order in QCD) and $3.78\pm 0.08\unit{pb}$ (at next-to-next-to-leading order in QCD), respectively~\cite{HiggsHandbook4}. As a benchmark, the branching fraction of $\Ph\to\ab\ab$ is assumed to be 10\%. The branching fractions of \ab to SM particles depend on the type of 2HDM+S, on the pseudoscalar mass \ma, and on \tanb, defined as the ratio of the vacuum expectation values of the second and first doublets. The $\tanb$ parameter is assumed to be $2$ which implies $2 \mathcal{B}(\ab \to \bbbar)\mathcal{B}(\ab \to \MM )=1.7\times10^{-3}$ for $\ma=30\GeV$ in type-III 2HDM+S~\cite{PhysRevD.90.075004}. For the set of parameters under discussion and with $\lbound\leq\ma\leq \hbound\GeV$, no strong dependence on $\ma$ is expected for $\mathcal{B}(\ab\to \bbbar)$ and $\mathcal{B}(\ab\to\MM)$~\cite{PhysRevD.90.075004}. The product of the cross section and branching fraction is therefore approximated to be about $8\unit{fb}$ for all $\ma$ values considered in this analysis.

The present search for the exotic $\ab$ particle in the $\mmbb$ final state is sensitive to the mass range of $\lbound\leq\ma\leq \hbound\GeV$. The sensitivity of the search largely decreases towards $\ma\approx20\GeV$ and lower because \ab gets boosted and the two \cPqb\ quark jets tend to merge~\cite{Curtin:2014pda}.
The upper bound is imposed by the Higgs boson mass. The analysis is performed using the proton-proton collision data at 13\TeV collected with the CMS detector during 2016, corresponding to an integrated luminosity of $\il$. Though the signal selection is optimised for the $\Ph\to\ab\ab\to\mmbb$ process, decays of ${\Ph}\to{\ab\ab}\to\mmtt$ can contribute to the selected sample if hadronically decaying $\Pgt$ leptons are misidentified as \cPqb\ quark jets. Such a contribution is found to be negligible using the benchmark scenario, although in some parts of the parameter space the enhancement in $\mathcal{B}(\ab\to\TT)$ can lead to a nonnegligible fraction of these events surviving the selection. This is taken into account in the scan over the ($\ma,\tanb$) plane in the type III 2HDM+S, as for certain values, the increase in $\mmtt$ signal can affect the sensitivity. The signal from $\ab\ab\to\bbtt$ with $\Pgt\to\Pgm$ leads to \mmumu significantly smaller than \ma and is not considered in the search.

The CMS detector is briefly described in Section~\ref{sect:cms}. The data and simulated samples are introduced in Section~\ref{sect:samples}. Section~\ref{sect:selection} is devoted to the event selection and categorisation. The signal and background modelling is discussed in Section~\ref{sec:background}, while in Section~\ref{sect:systematics}, different sources of systematic uncertainties are described. Results are presented in Section~\ref{sec:results}, and the paper is summarised in Section~\ref{sec:summary}.
\section{The CMS detector}
\label{sect:cms}
The central feature of the CMS apparatus is a superconducting solenoid of 6\unit{m} internal diameter, providing a magnetic field of 3.8\unit{T}. Within the solenoid volume are a silicon pixel and strip tracker, a lead tungstate crystal electromagnetic calorimeter, and a brass and scintillator hadron calorimeter, each composed of a barrel and endcap sections. Forward calorimeters, made of steel and quartz-fibres, extend the pseudorapidity coverage provided by the barrel and endcap detectors. Muons are detected in gas-ionisation chambers embedded in the steel flux-return yoke outside the solenoid.
They are measured in the pseudorapidity range $\abs{\eta} < 2.4$, with detection planes made using three technologies: drift tubes, cathode strip chambers, and resistive plate chambers. The efficiency to reconstruct and identify muons is greater than 96\%. Matching muons to tracks measured in the silicon tracker results in a relative transverse momentum (\pt) resolution, for muons with \pt up to 100\GeV, of 1\% in the barrel and 3\% in the endcaps~\cite{Sirunyan:2018}.
A more detailed description of the CMS detector, together with a definition of the coordinate system used and the relevant kinematic variables, can be found in Ref.~\cite{Chatrchyan:2008zzk}.
\section{Simulated samples}
\label{sect:samples}
The NMSSMHET model~\cite{PhysRevD.90.075004} is used to generate signal samples with the Monte Carlo (MC) event generator {\MGvATNLO}~\cite{Alwall:2014hca} at leading order (LO). The signal samples span the \ma search region in 5\GeV steps. Background processes with dominant contributions are the Drell--Yan production in association with additional \cPqb\ quarks and \ttbar in the dimuon final state. Simulated samples for background processes are used in this analysis to optimise the selection and for validation purposes in those selection steps that yield reasonable statistical precision. The contribution of backgrounds to the selected sample is directly extracted from data with no reference to simulation. The Drell--Yan process, ${\PZ}/\cPgg^*(\to\ell^+\ell^-)+\,\text{jets}$ with a minimum dilepton mass threshold of 10\GeV, is modelled with the same event generator at LO, exclusive in number of additional partons (up to 4). The reference cross section for the Drell--Yan process is computed using {\FEWZ}\,3.1~\cite{Li:2012wna} at next-to-next-to-leading order.
The top quark samples, \ttbar and single top quark production, are produced with {\POWHEG}2.0~\cite{Re:2010bp,Alioli:2010xd,Alioli:2009je,Frixione:2007vw} at next-to-leading order (NLO). Backgrounds from diboson ($\PW\PW$, $\PW\PZ$, $\PZ\PZ$) production are generated at NLO with the same program and settings as that of the Drell--Yan samples. The only exception is the \PW\PW\ process that is generated at LO. The set of parton distribution functions (PDFs) is NLO NNPDF3.0 for NLO samples, and LO NNPDF3.0 for LO samples~\cite{Ball:2014uwa}. For all samples, \PYTHIA\,8.212~\cite{Sjostrand:2014zea} with tune CUETP8M1~\cite{Skands:2014pea,Khachatryan:2015pea} is used for the modelling of the parton showering and fragmentation. The full CMS detector simulation based on {\GEANTfour}~\cite{geant4} is implemented for all generated event samples. In order to model the effect of
additional interactions per bunch crossing (pileup), generated minimum bias events are added to the simulated samples. The number of additional interactions are scaled to agree with that observed in data~\cite{CMS-PAS-LUM-17-001}.
\section{Event selection and categorisation}
\label{sect:selection}
Events are filtered using a high-level trigger requirement based on the presence of two muons with $\pt>17$ and $8\GeV$. For offline selection, events must contain at least one primary vertex, considered as the vertex of the hard interaction. At least four tracks must be associated with the selected primary vertex. The longitudinal and radial distances of the vertex from the centre of the detector must be smaller than 24 and 2\unit{cm}, respectively. The vertex with the largest sum of $\pt^{2}$ of the physics objects is chosen for the analysis. The physics objects are the jets, clustered using the jet finding algorithm~\cite{antikt,Cacciari:2011ma} with the tracks assigned to the vertex as inputs, and the associated missing transverse momentum, taken as the negative vector sum of the \pt of those jets. Extra selection criteria are applied to leptons and jets, reconstructed using the CMS particle-flow algorithm~\cite{Sirunyan:2017ulk}.

The selection requires two muons with opposite electric charge in $\abs{\eta}<2.4$, originating from the selected primary vertex. Events with the leading (subleading) muon $\pt >20\,(9)\GeV$ are selected.
A relative isolation variable $\PFrelIso$ is calculated by summing the transverse energy deposited by other particles inside a cone of size $\Delta R = \sqrt{\smash[b]{(\Delta\eta)^2+(\Delta\phi)^2}}=0.4$ around the muon with $\phi$ being the azimuthal angle measured in radians, divided by the muon $\pt$,
\begin{linenomath}
\begin{equation}
\PFrelIso = \frac{ \pfChargedHadronIso + \text{max}((\pfPhotonIso + \pfNeutralHadronIso - 0.5\,\pfPU),0)}{\pt},
\label{eq:pfreliso}
\end{equation}
\end{linenomath}
where $\pfChargedHadronIso$, $\pfPhotonIso$, $\pfNeutralHadronIso$ and \pfPU\ are, respectively, the scalar \pt sums of stable charged hadrons, photons, neutral hadrons, and charged hadrons associated with pileup vertices. The contribution $0.5\,\pfPU$ accounts for the expected pileup contribution from neutral particles. The neutral-to-charged particle ratio is taken to be approximately 0.5 from isospin invariance. Only muons with the isolation variable satisfying $\PFrelIso<0.15$ are considered in the analysis. The efficiencies for muon trigger, reconstruction, and selection in simulated events are corrected to match those in data. In case more muons in the event pass the selection requirements, the two with the largest $\pt$ are chosen.

Jets are reconstructed by clustering charged and neutral particles using the anti-$\kt$ algorithm~\cite{antikt} with a distance parameter of 0.4. The reconstructed jet energy is corrected for effects from the detector response as a function of the jet $\pt$ and $\eta$. Contamination from pileup, underlying event, and electronic noise are subtracted~\cite{CMS-PAS-JME-16-003,CMSJetPaper}. Extra $\eta$-dependent smearing is performed on the jet energy in simulated events as prescribed in Refs.~\cite{CMS-PAS-JME-16-003,CMSJetPaper}.

Events are required to have at least two jets with $\abs{\eta}<2.4$ and $\pt>20$ (leading) and 15\GeV (subleading), with both jets separated from the selected muons ($\Delta R > 0.5$). A combined secondary vertex algorithm is used to identify jets that are likely to originate from \cPqb\ quarks. The algorithm uses the track-based lifetime information together with the secondary vertices inside the jet to provide a multivariate discriminator for the \cPqb\ jet identification~\cite{Sirunyan:2017ezt}.
Working points ``loose'' (L), ``medium'' (M), and ``tight'' (T) are defined. They correspond to thresholds on the discriminator, for which the misidentification probability is around 10, 1, and 0.1\%, respectively, for jets originating from light quarks and gluons~\cite{Sirunyan:2017ezt}. The misidentification probability for jets originating from \cPqc\ quarks is around 30, 10, and 2\%, respectively, for the loose, medium, and tight working points. The efficiencies for correctly identifying \cPqb\ jets are $\approx$80\% for the loose, $\approx$60\% for the medium, and $\approx$40\% for the tight working point. The jet with maximum discriminator value must pass the tight working point of the algorithm, while the second is required to pass the loose one. The correction factors for \cPqb\ jet identification are applied to simulated events to reproduce the data distribution of the \cPqb\ tagging discriminator. In events with more jets passing the selection criteria, the two with the largest $\pt$ are taken.

The imbalance in the transverse momentum in signal events is not expected to be large, as the contribution from neutrinos from semileptonic decays in \cPqb\ jets is typically small. The missing transverse momentum, $\ptmiss$ is defined as the magnitude of the negative vector sum of the transverse momenta of all reconstructed particles.
The jet energy calibration introduces corrections to the $\ptmiss$ measurement~\cite{CMSJetPaper}. Events are required to have $\ptmiss<60\GeV$.

\begin{figure}[!htb]
\centering
	\includegraphics[width=0.48\textwidth]{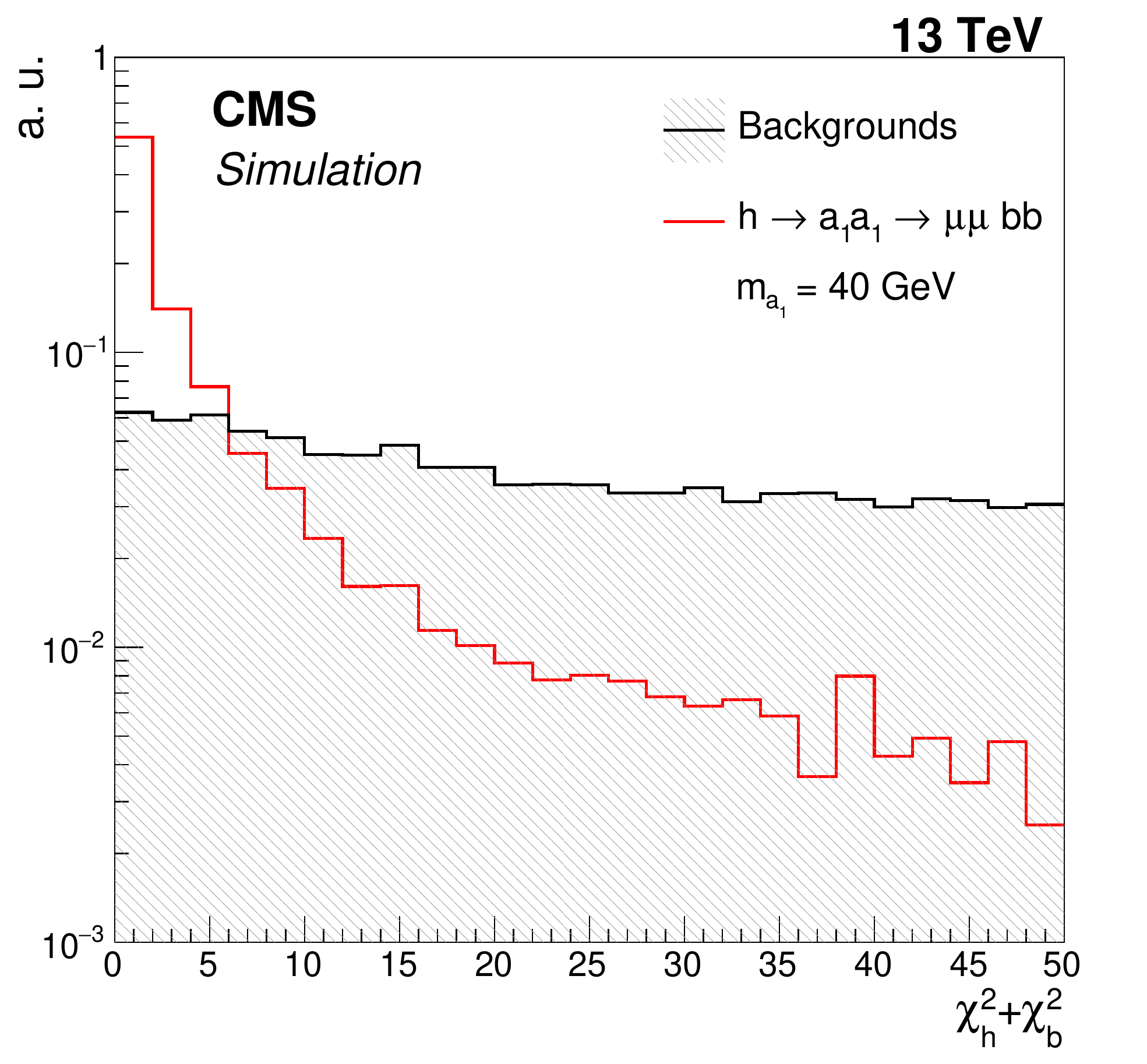}
  \caption{The distribution of $\chit$ in simulated background processes and the signal process with $\ma=40\GeV$. The samples are normalised to unity. }
  \label{fig:p0}
\end{figure}

\begin{figure}[!htb]
\centering
	\includegraphics[width=0.48\textwidth]{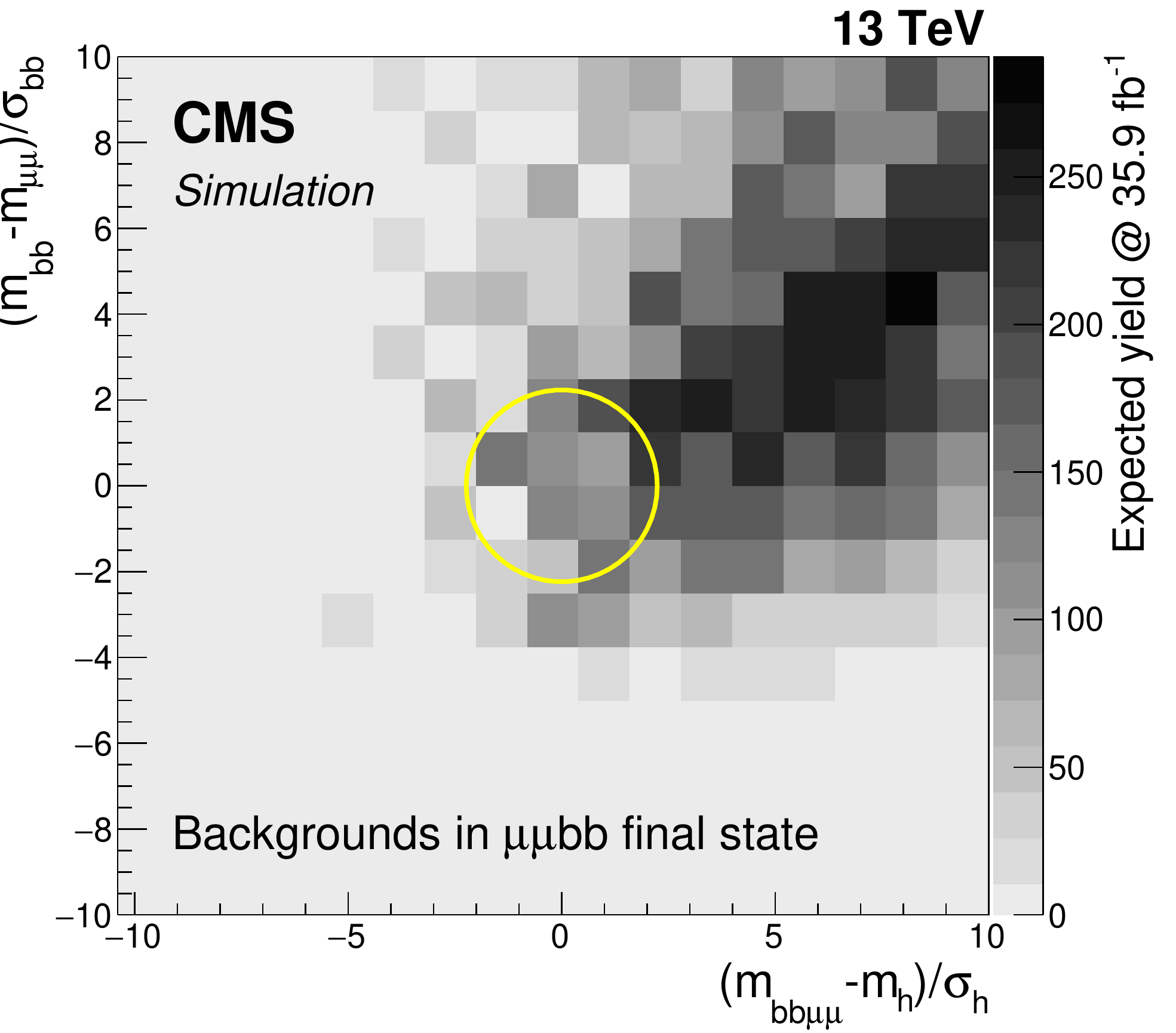}
	\includegraphics[width=0.48\textwidth]{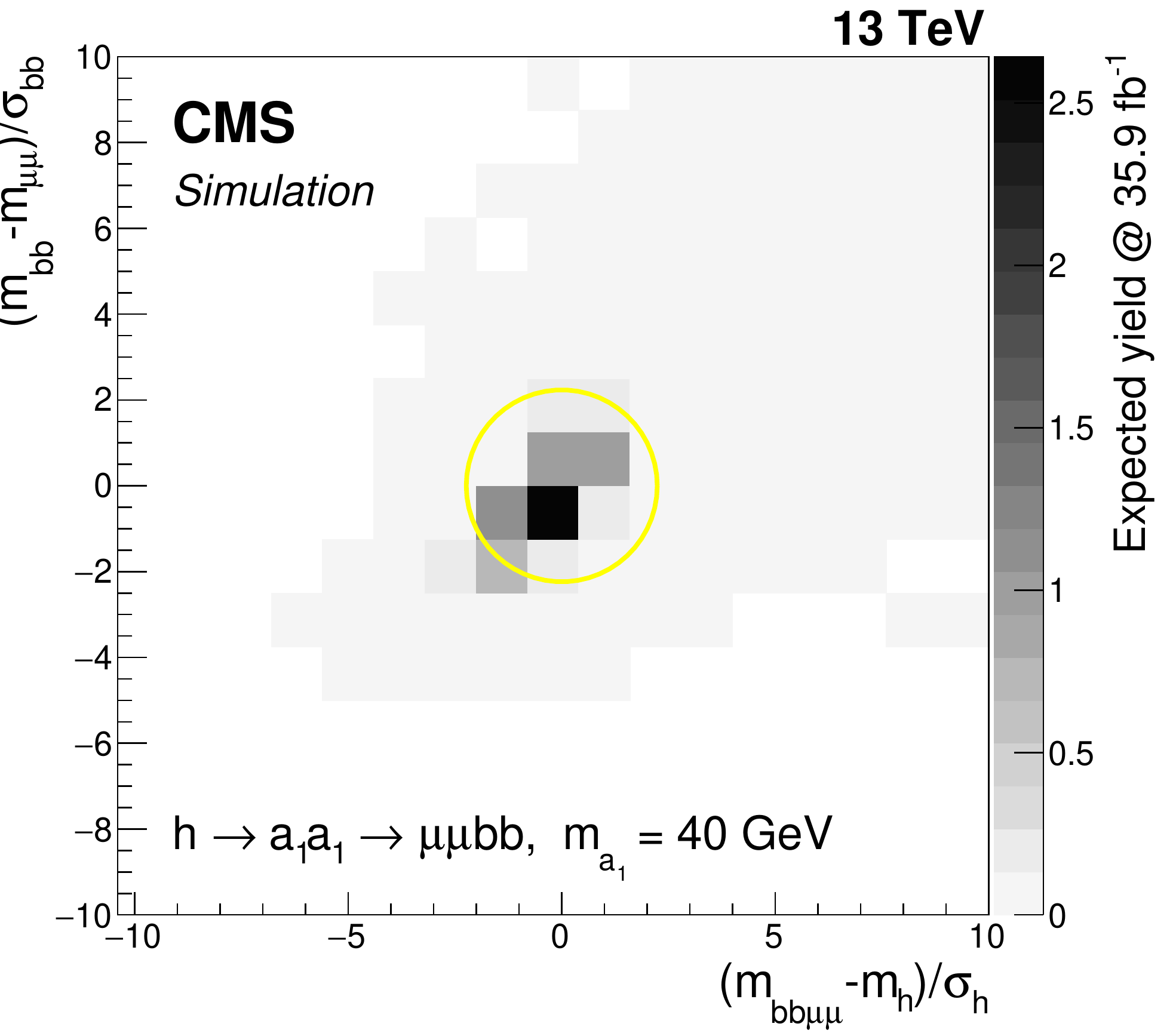} \\	
  \caption{The distribution of $\chib$ versus $\chih$ as defined in Eq.~(\ref{eq:chi2}) for (\cmsLeft) simulated background processes and (\cmsRight) the signal process with $\ma=40\GeV$. The contours encircle the area with $\chit<5$. The grey scale represents the expected yields at \il.}
  \label{fig:2Dchi}
\end{figure}

Assuming the \cPqb\ quark jets and muons are the decay products of the pseudoscalar \ab, it is expected to have $\mbb{\approx}\mmumu{\approx}\ma$ in signal events. Moreover, the system of muons and \cPqb\ quark jets is expected to have an invariant mass close to $m_{\Ph}$. A $\chit$ variable is introduced as $\chib^2+\chih^2$, where
\begin{equation}
  \label{eq:chi2}
 \chib= \frac{(\mbb - \mmumu)}{\sigma_{\cPqb\cPqb}}\qquad\text{and}\qquad\chih=\frac{(\mfour - m_{\Ph})}{\sigma_{\Ph}}.
\end{equation}
Here $\sigma_{\cPqb\cPqb}$ and $\sigma_{\Ph}$ are, respectively, the mass resolutions of the di-\cPqb-quark jet system and the Higgs boson candidate, derived from simulation. The mass resolution of the di-\cPqb-quark jet system increases linearly with $\ma$. It is evaluated on an event-by-event basis, where $\mmumu$ is assumed to be equal to $\ma$. The decay width of \ab is negligible compared with the experimental mass resolutions in the analysis. The distribution of \chit in the signal sample with $\ma=40\GeV$ is compared with that in backgrounds in Fig.~\ref{fig:p0}. Events are selected with $\chit<5$. In Fig.~\ref{fig:2Dchi}, \chib and \chih are shown in 2D histograms for backgrounds and for the signal with $\ma=40\GeV$, where the contour of $\chit<5$ is also presented. This selection has a signal efficiency up to 64\% while rejecting more than 95\% of backgrounds. The tails in the \chib and \chih distributions, arising from the imperfect energy estimation of \cPqb\ jets as well as combinatorics of the di-\cPqb-jet system, are more populated in background processes. 
The search for the new particle \ab is performed within $\lbound\leq\ma\leq \hbound\GeV$. A slightly wider range, driven by the narrow width of \ab and the high resolution of \mmumu, is used for the event selection; thus events with \mmumu values not in $[19.5,63.5]\GeV$ are discarded. This ensures the full signal selection efficiency and the proper background modelling at the boundaries.

\begin{figure*}[!htbp]
\centering
	\includegraphics[width=0.48\textwidth]{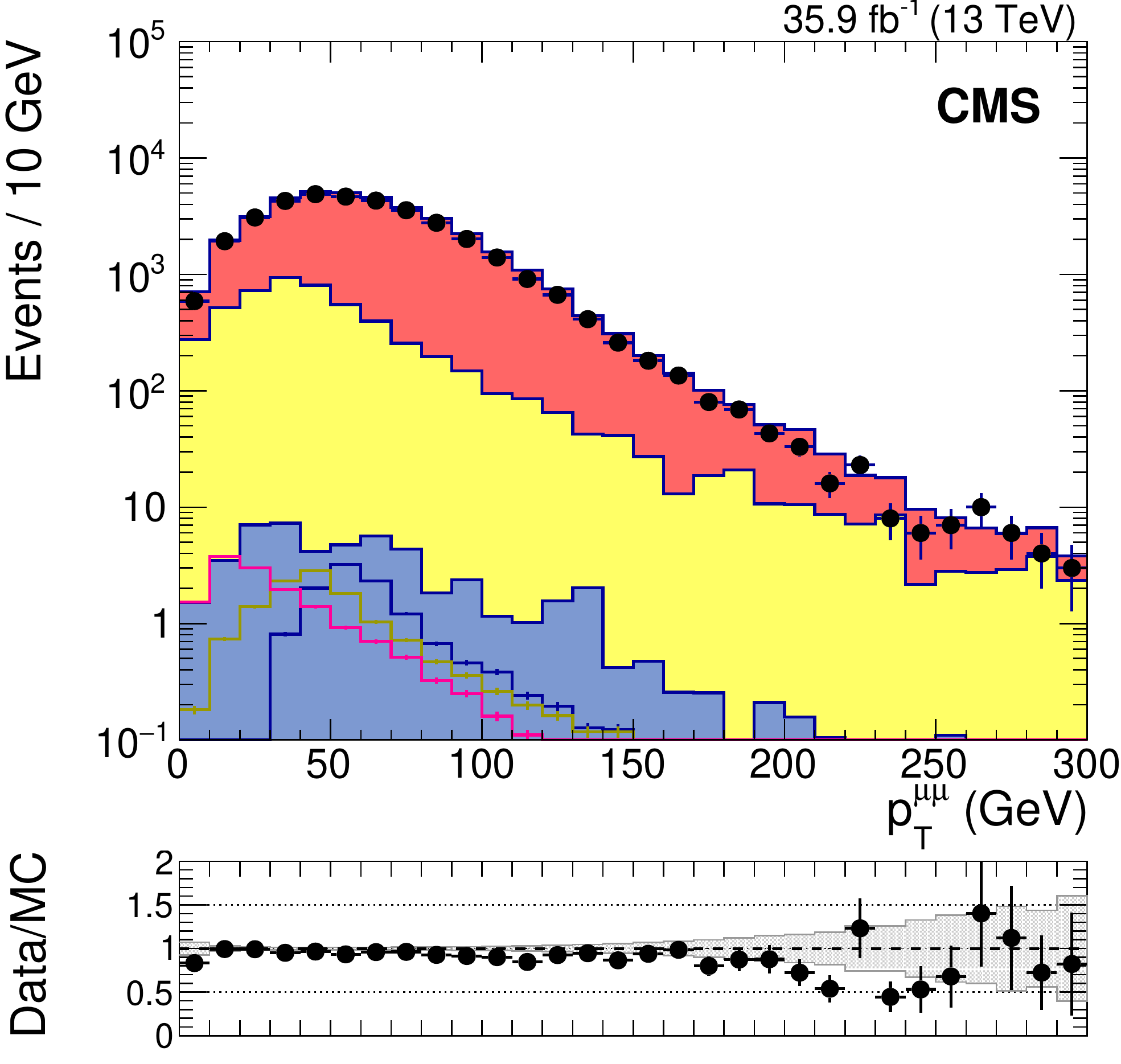}
	\includegraphics[width=0.48\textwidth]{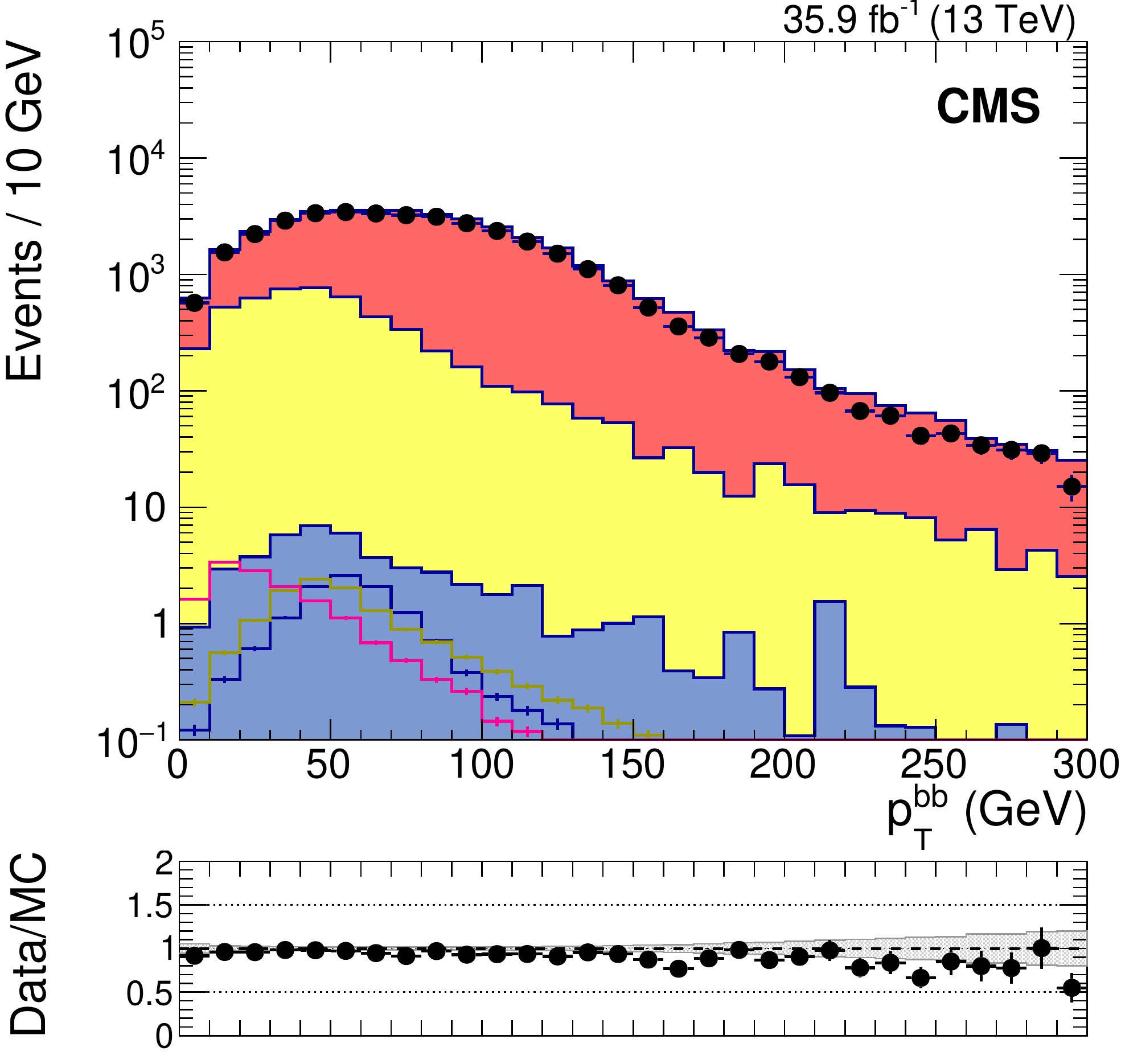} \\	
	\includegraphics[width=0.48\textwidth]{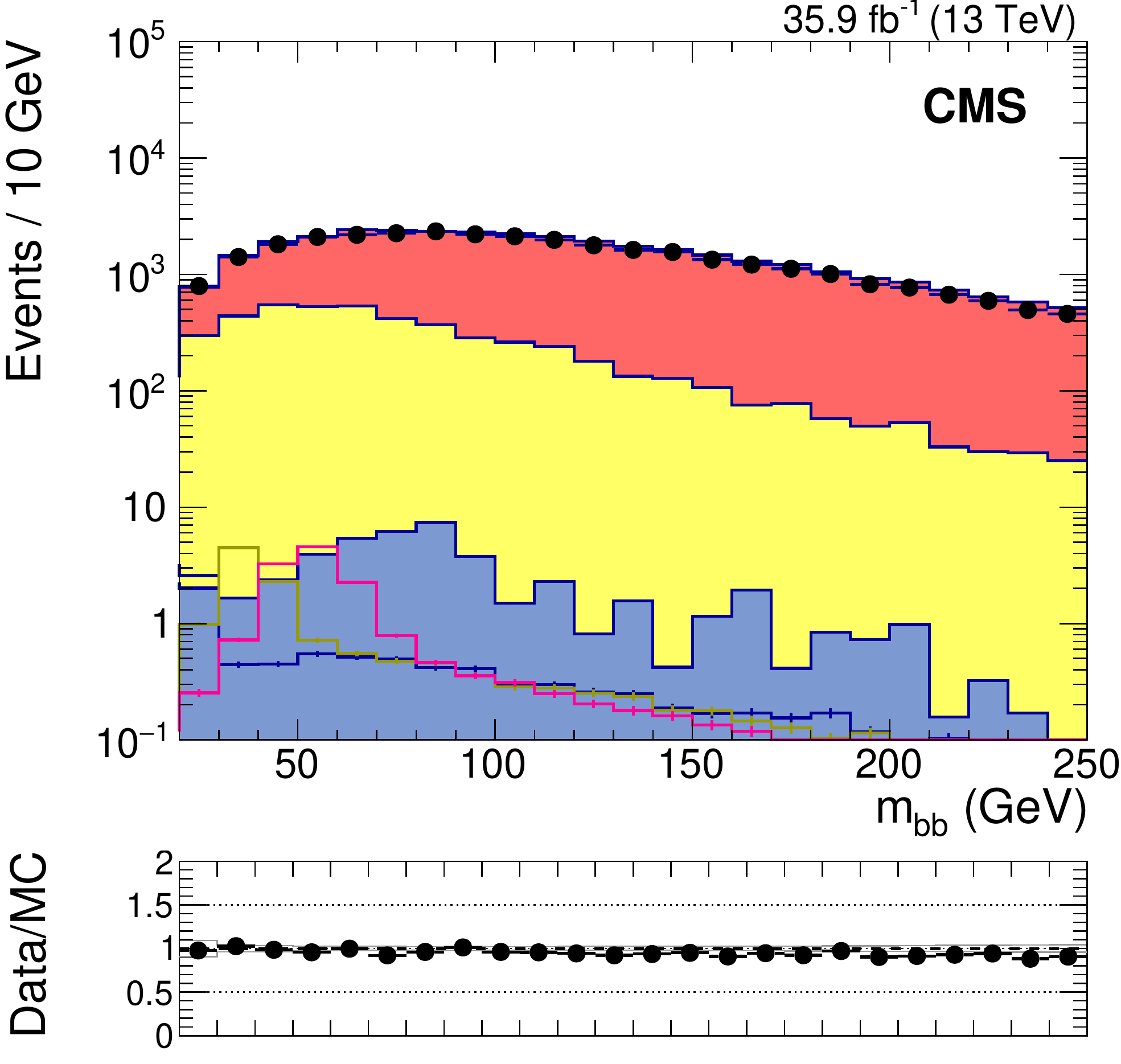}
	\includegraphics[width=0.48\textwidth]{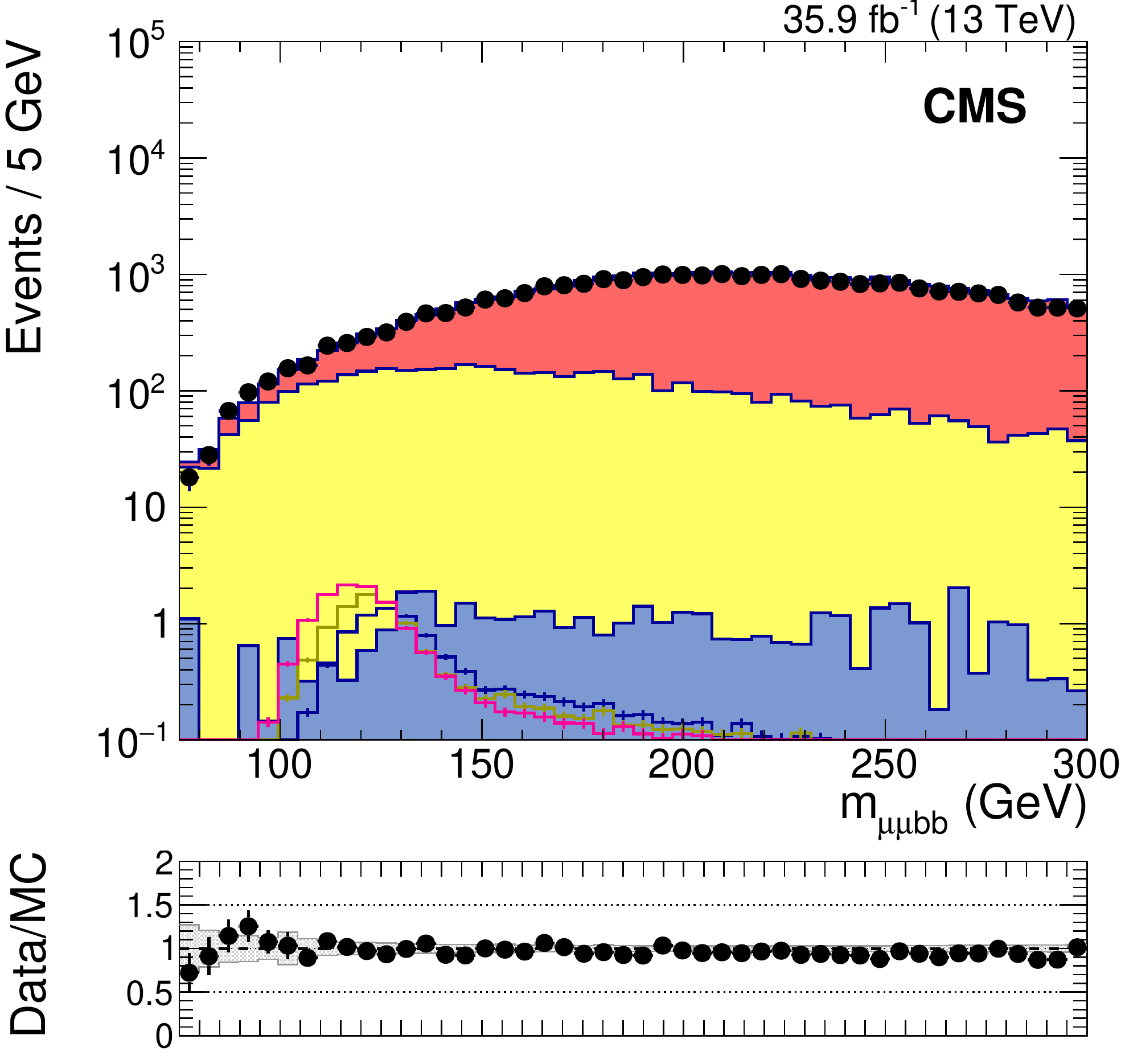} \\
	\includegraphics[width=0.48\textwidth]{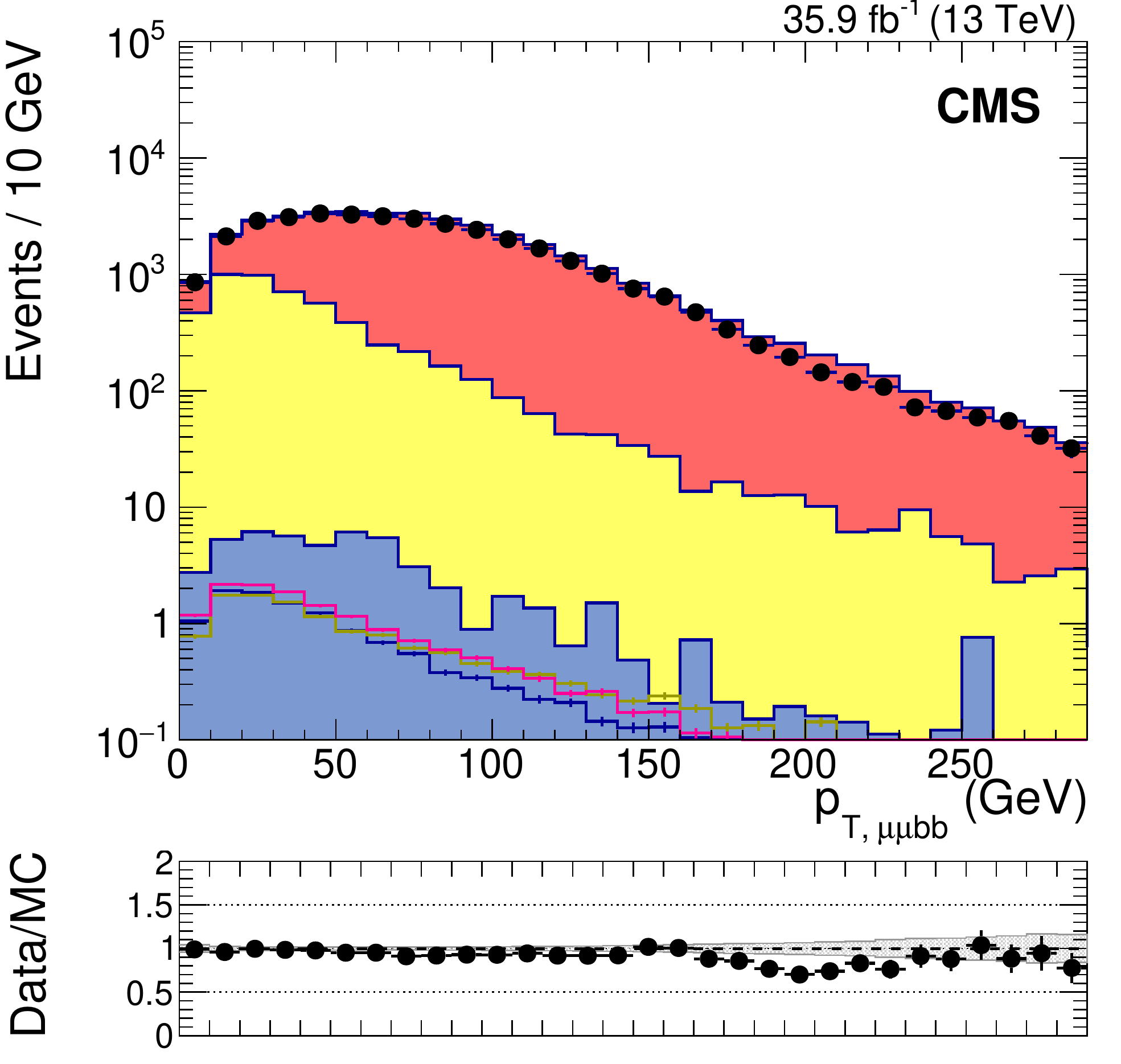}
	\includegraphics[width=0.48\textwidth]{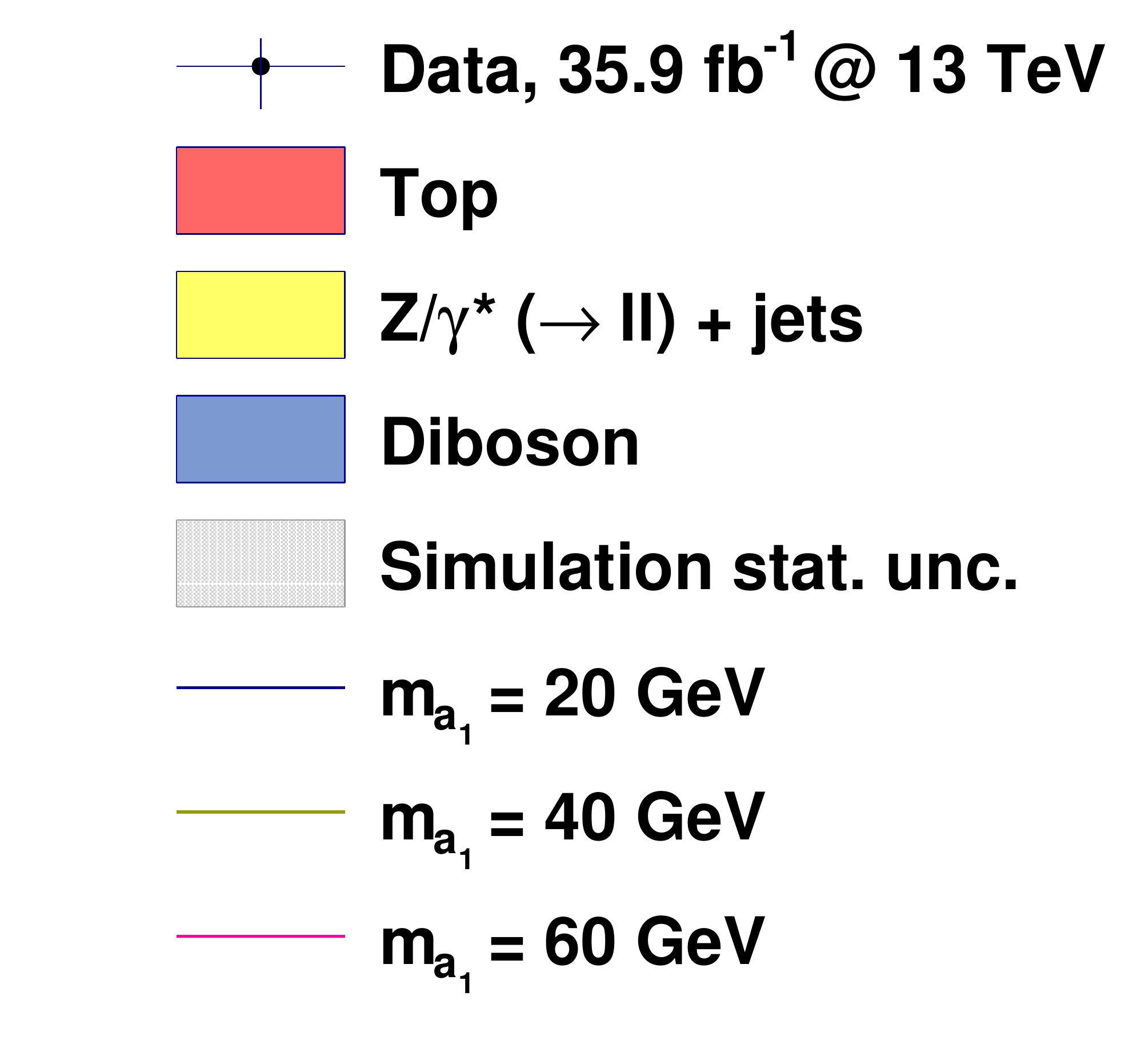} \\	
  \caption{The distribution of the $\pt$ of the (top left) dimuon and (top right) di-\cPqb-jet system, the mass of the (middle left) di-\cPqb-jet and (middle right) $\Pgm\Pgm {\cPqb\cPqb}$ system, and (bottom left) the $\pt$ of the $\Pgm\Pgm {\cPqb\cPqb}$ system, all after requiring two muons and two \cPqb-tagged jets in the event. Simulated samples are normalised to an integrated luminosity of $\il$ using their theoretical cross sections.}
  \label{fig:p1}
\end{figure*}

A method that fully relies on data is used to estimate the background, as described in Section~\ref{sec:background}. Simulated background samples are however used to optimise the selection. Figure~\ref{fig:p1} shows distributions, in data and simulation, for events passing the selection requirements except those of $\ptmiss$ and $\chit$. In this figure, expected number of simulated events is normalised to the integrated luminosity of $\il$. Data and simulation are compared for the $\pt$ of the dimuon system, and the mass and $\pt$ of the di-\cPqb-jet system. Using the same selected muon and jet pairs, Fig.~\ref{fig:p1} also illustrates the distributions of the invariant mass $\mfour$ and the transverse momentum $\pt^{\Pgm\Pgm\cPqb\cPqb}$ of the four-body system. The distributions for simulated events follow reasonably those in the data, within the statistical uncertainties presented in the figure. The yield in data and the expected yields in simulation are presented in Table~\ref{tab:selection}. The expected yield from a signal of $\Ph\to{\ab\ab}\to\mmtt$ is found to be around 0.01 with the model parameters used in this table.

\begin{table*}[!ht]
\topcaption{Event yields for  simulated processes and data after requiring two muons and two \cPqb\ jets (\mmbb selection) and after the final selection. The expected number of simulated events is normalised to the integrated luminosity of $\il$. Uncertainties are statistical only.}
\centering
    \begin{tabular}{lcc}

			Process &	\mmbb selection & Final selection \\	
\hline

	Top (\ttbar, single top quark)	        &$33\,730 \pm 120$	&$198 \pm 9$		\\

	Drell--Yan 				&	$5237 \pm 77$	&	$399 \pm 21$	\\

	Diboson					&$51 \pm 4$		&	$1 \pm 0.1$	\\

	Total expected background		&$39\,015 \pm 140$	&$598 \pm 23$	\\

	Data							&36\,360	&610			\\

Signal for $\sigma_{\Ph}\mathcal{B} \approx 8\unit{fb}$ &&		\\
~~~$\ma = 20\GeV$& $14.0\pm 0.1$&$6.0 \pm 0.1$ \\
~~~$\ma = 40\GeV$& $14.8\pm 0.1$&$7.5 \pm 0.1$ \\
~~~$\ma = 60\GeV$& $16.7\pm 0.1$&$10.1 \pm 0.1$ \\

   \end{tabular}
\label{tab:selection}
\end{table*}

To enhance the sensitivity, an event categorisation is employed: different categorisation schemes are tried, and the one resulting in the highest expected significance is chosen. The data in a sideband region are used to determine the categorisation
that is most sensitive for this analysis. The sideband region is constructed using the same selection as that for the signal region except that $5<\chit<11$.
In simulated background samples, the correlations between $\chit$ and $\mmumu$ and the variables used for categorisation are found to be small. The best sensitivity is found with categorisation according to the \cPqb\ tagging discriminator value of the loose \cPqb-tagged jet. The tight-tight (TT) category contains events with both jets passing the tight requirements of the \cPqb\ jet identification algorithm. Events in which the loose \cPqb-tagged jet passes the medium \cPqb\ tagging requirements but fails the tight conditions fall into the tight-medium (TM) category. The remaining events with the loose \cPqb-tagged jet failing the medium requirements of the \cPqb\ jet identification algorithm belong to the tight-loose (TL) category. On average, 41\% of signal events pass the TL selection, while 32\% fulfil the TM requirements and 27\% belong to the TT category. According to the data in the sideband region, the majority of background events ($\approx$70\%) fall into the TL category whereas about 20\% pass the TM requirements and less than 10\% can meet the TT criteria.
\section{Signal and background modelling}
\label{sec:background}
The search is performed using an unbinned fit to the $\mmumu$ distribution in data, simultaneously in the TT, TM, and TL categories. The signal shape is modelled with a weighted sum of Voigt profile~\cite{2008arXiv0805.2274P} and Crystal Ball (CB) functions~\cite{CrystalBallRef}, where the mean values of the two are bound to be the same. The initial values for the signal model parameters are extracted from a simultaneous fit of the model to the simulated signal samples described in Section~\ref{sect:samples}. Almost all parameters in the signal model are found to be independent of $\ma$ and are fixed in the final fit. The only exceptions are the resolution parameter of the Voigt profile and CB functions, $\sigma_{\mathrm{v}}$ and $\sigma_{\mathrm{cb}}$, respectively. These parameters depend linearly on \ma and only their slopes, respectively $\alpha$ and $\beta$, float in the final fit within their uncertainties,
\begin{linenomath}
\begin{equation}
\label{eqn:sigparam}
  \begin{aligned}
\sigma_{\mathrm{v}} &=& \sigma_{\mathrm{v},0}  + \alpha\,\mmumu,\\
\sigma_{\mathrm{cb}} &=& \sigma_{\mathrm{cb},0}  + \beta\,\mmumu.
  \end{aligned}
\end{equation}
\end{linenomath}
The expected signal efficiency and acceptance are interpolated for \ma values not covered by simulation. The \mmumu distribution in data is used to evaluate the contribution of backgrounds. The uncertainty associated with the choice of the background model is treated in a similar way as other uncertainties for which there are nuisance parameters in the fit. The unbinned likelihood function for the signal-plus-background fit has the form
\begin{linenomath}
\begin{equation}
\label{eq:ll1}
\mathcal{L}({\mathrm{data}}|s(p,\mmumu)+b(\mmumu)),
\end{equation}
\end{linenomath}
where $s(p,\mmumu)$ is the parametric signal shape with the set of parameters indicated by $p$, and $b(\mmumu)$ is the background model. The shape for the background is modelled, independently in each category, with a set of analytic functions using the discrete profiling method~\cite{Dauncey:2014xga,Khachatryan:2014ira,Aad:2015zhl}. In this approach the choice of the functional form of the background shape is considered as a discrete nuisance parameter, for which the best fit value can vary as the trial value of the parameter of interest (\mmumu) varies. The background parameter space therefore contains multiple models, each including its own parameters.

To provide the input background models to the discrete profiling method, the data are modelled with different parametrisations of polynomials.
The degrees of the polynomials are determined through statistical tests (F-test)~\cite{James:2006zz} to ensure the sufficiency of number of parameters and to avoid over-fitting the data.
The input background functions are tried in the minimisation of the negative logarithm of the likelihood with a penalty term added to account for the number of free parameters in the background model. The discrete profiling method can choose a different best-fit functional form for the background as the physics parameter of interest varies, thus effectively incorporating
the systematic uncertainty on the background functional form: in the present analysis the result is to yield expected upper limits that are about 10\% less stringent than those obtained with a single functional form for the background. The likelihood ratio for the penalised likelihood function $\widetilde{\mathcal{L}}$ can be written as
\begin{linenomath}
\begin{equation}
\label{eq:ll}
-2 \ln \frac{\widetilde{\mathcal{L}}({\mathrm{data}}|\mu, \hat\theta_\mu, \hat b_\mu)}{\widetilde{\mathcal{L}}({\mathrm{data}}|\hat\mu, \hat\theta, \hat b)},
\end{equation}
\end{linenomath}
where $\mu$ is the measured quantity. The numerator is the maximum penalised likelihood for a given $\mu$, at the best fit values of nuisance parameters, $\hat\theta_\mu$ and of the background function, $ \hat b_\mu$. The denominator is the global maximum for $\widetilde{\mathcal{L}}$, achieved at $\mu=\hat\mu$, $\theta = \hat\theta$ and $b = \hat b$. A confidence interval for $\mu$ is obtained with the background function maximising $\widetilde{\mathcal{L}}$ for any value of $\mu$~\cite{Dauncey:2014xga}.
\section{Systematic uncertainties}
\label{sect:systematics}
The statistical interpretation of the analysis takes into account several sources of systematic uncertainties related to the accuracy in the signal modelling and uncertainties in the signal acceptance. The imprecise knowledge of the background contributions is taken into account by the discrete profiling method described in Section~\ref{sec:background}.

\textit{Theoretical uncertainties:} to evaluate the upper limit on $\mybr$, the Higgs boson production cross section is set to the SM prediction where an uncertainty of 4.7\% is considered for the sum of the $\cPg\cPg$F and VBF production cross sections, accounting for QCD scale, PDF, and $\alpha_s$ uncertainties~\cite{HiggsHandbook4}.

\textit{Uncertainties in signal shape and acceptance modelling:} an uncertainty of 2.5\% is assigned to the integrated luminosity of the CMS 13\,TeV data collected in 2016~\cite{CMS-PAS-LUM-17-001}. The uncertainty in the number of pileup interactions per event is estimated by varying the total inelastic $\Pp\Pp$ cross section by $\pm$4.6\%~\cite{Sirunyan:2018nqx}. The simulation-to-data correction factors for the trigger efficiency, muon reconstruction, and selection efficiencies are estimated using a ``tag-and-probe'' method~\cite{tagandprobe} in Drell--Yan data and simulated samples. These uncertainties include the pileup dependence of the correction factors. For the jet energy scale (JES), the variations are made according to the $\eta$- and $\pt$-dependent uncertainties and propagated to the \ptmiss of the event. An additional uncertainty, arising from unclustered energies in the event, is assessed for \ptmiss. For the jet energy resolution, the smearing corrections are varied within their uncertainties~\cite{CMS-PAS-JME-16-003}. Systematic uncertainty sources that affect the simulation-to-data corrections of the \cPqb\ tagging discriminator distribution are JES, the contaminations from light flavor (LF) jets in the \cPqb-jet sample, the contaminations from heavy flavor (HF) jets in the light-flavor jet sample, and the statistical fluctuations in data and MC. The uncertainties due to JES and light-flavor jet contamination in \cPqb-jet samples are found to be dominant~\cite{Sirunyan:2017ezt}. Uncertainties in the choice of the renormalisation, $\mur$, and factorisation, $\muf$, scales are estimated by doubling and halving $\mur$ and $\muf$ simultaneously in the signal sample. To estimate the uncertainties associated with the parton showering and fragmentation model, additional signal samples are produced using \HERWIG++~\cite{Bahr:2008pv} and compared to \PYTHIA. Finally, uncertainties arising from the limited understanding of the PDFs~\cite{Butterworth:2015oua} are taken into account. These uncertainties have a negligible effect on the shape of the signal. Their effects on the yield are taken into account by introducing nuisance parameters with log-normal distributions into the fit.
\section{Results}
\label{sec:results}
\begin{figure*}[!ht]
\centering
   \includegraphics[width=0.48\textwidth]{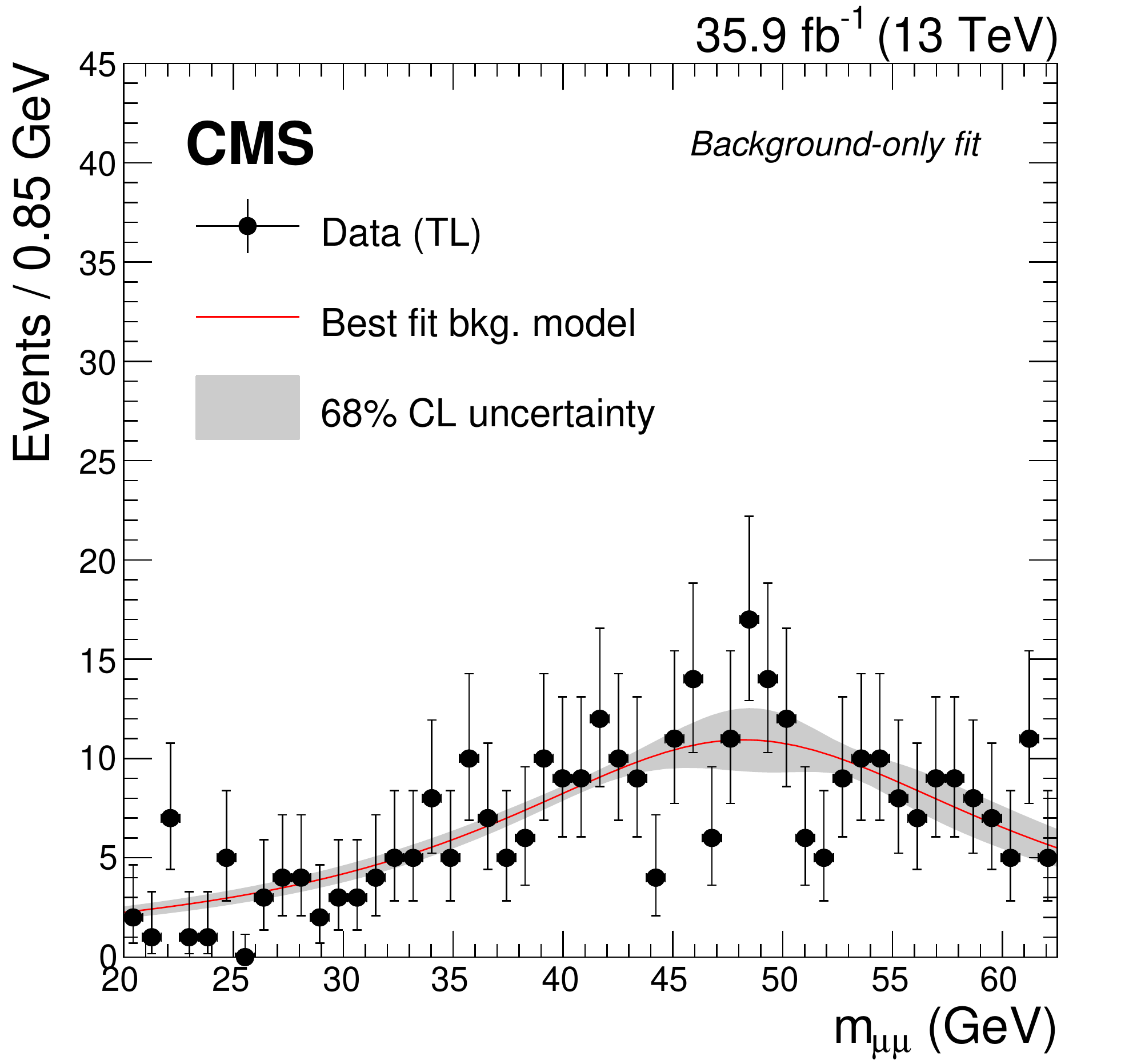}
   \includegraphics[width=0.48\textwidth]{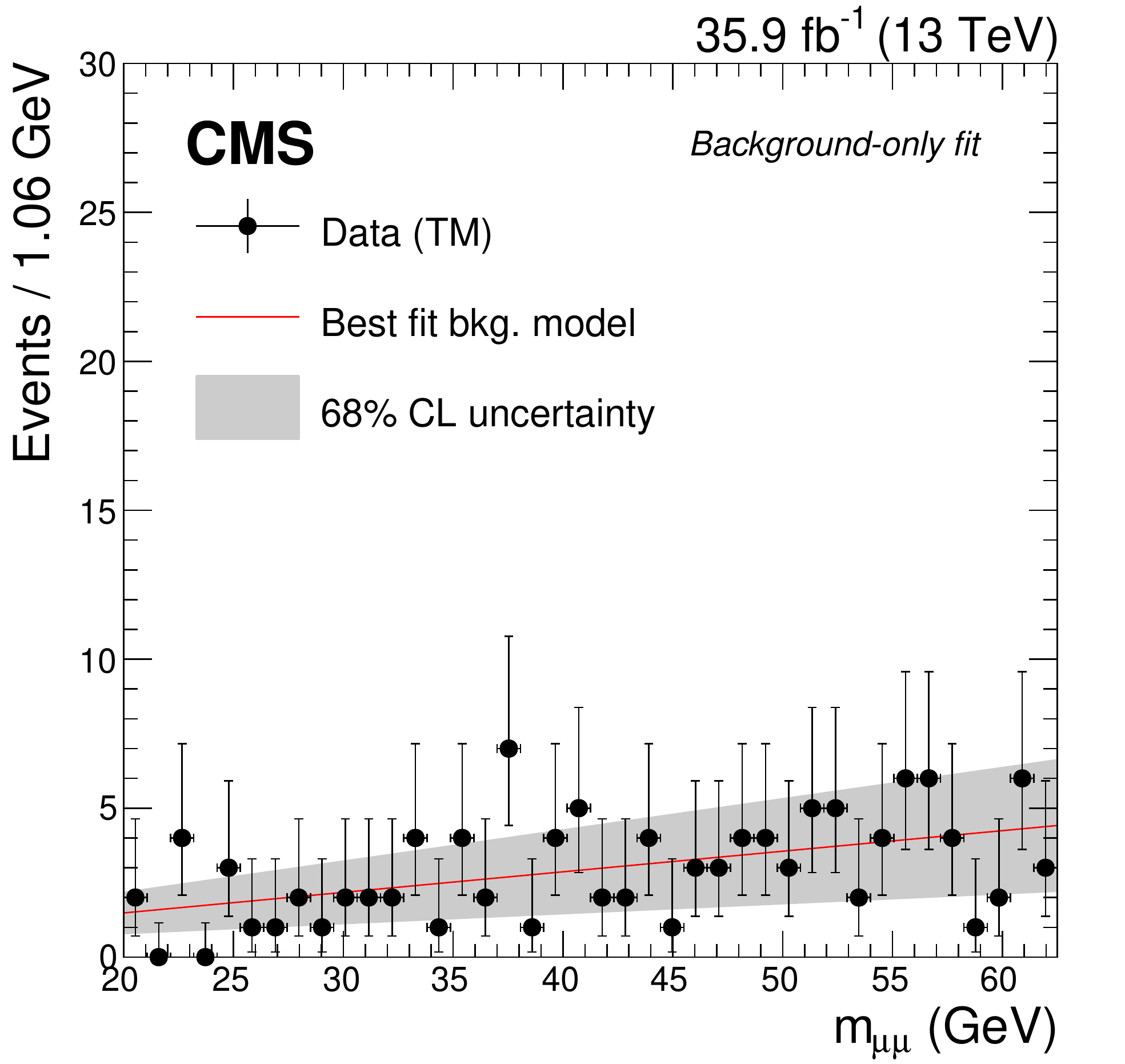}\\
   \includegraphics[width=0.48\textwidth]{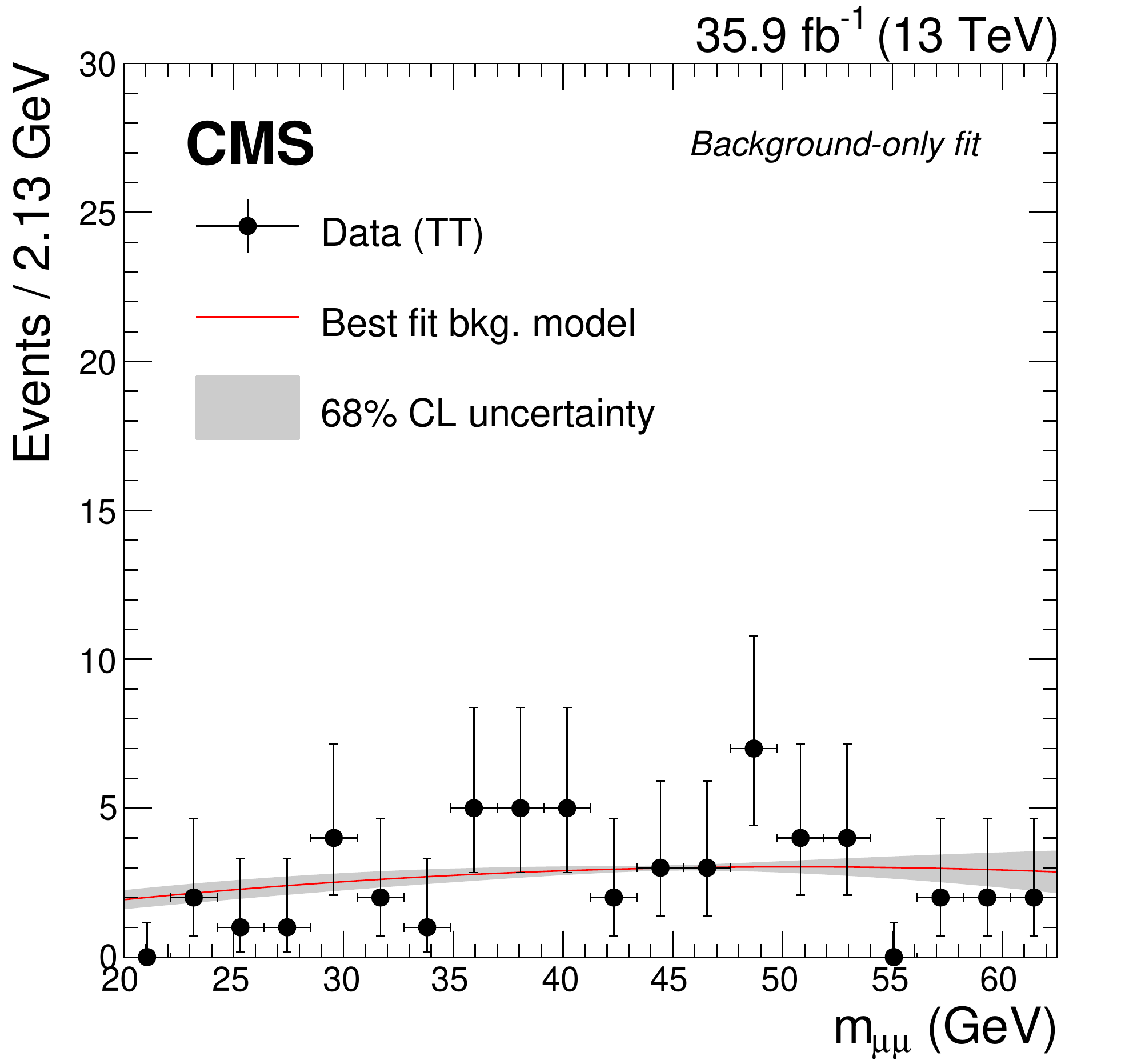}
   \includegraphics[width=0.48\textwidth]{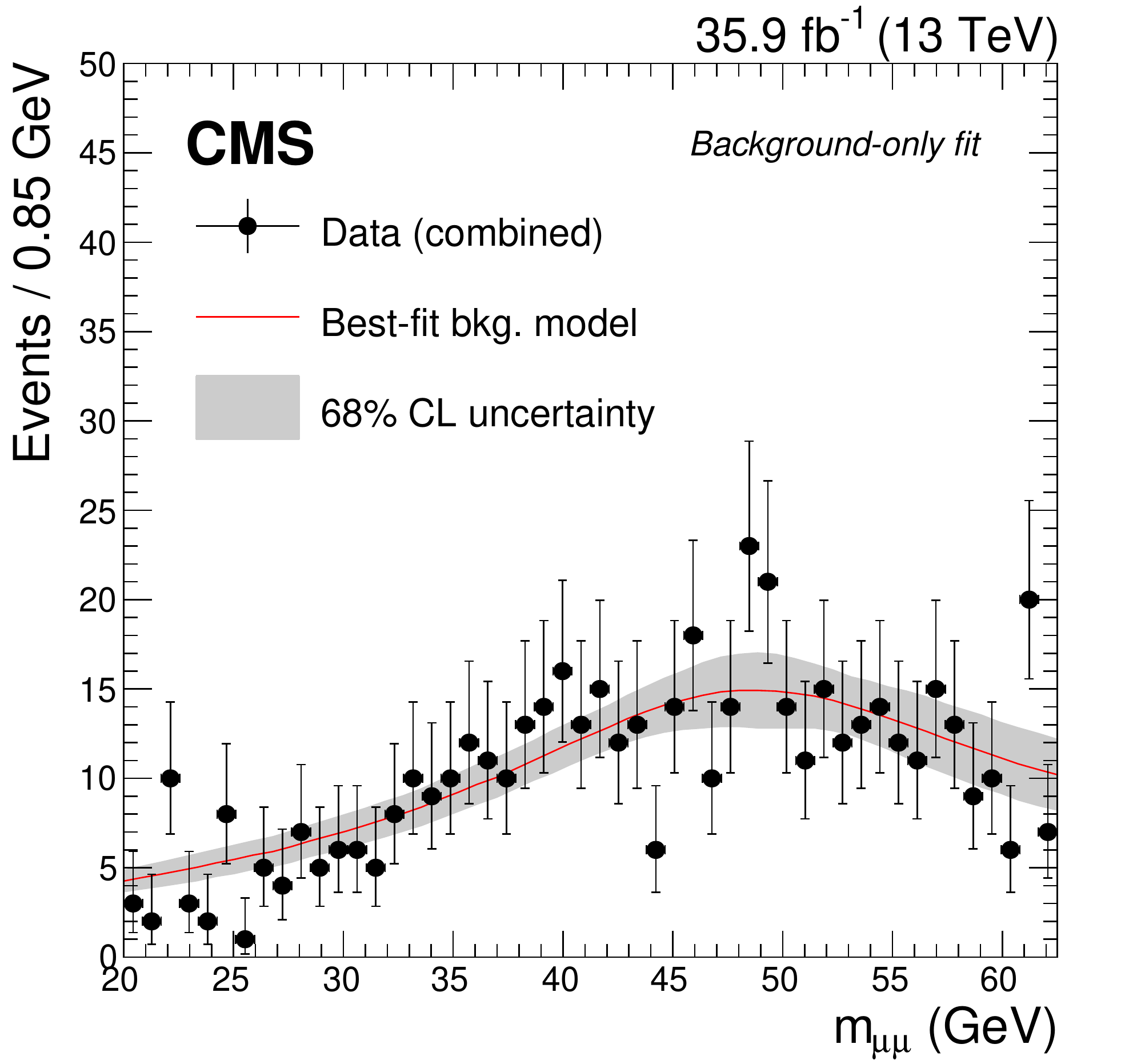}\\
   \caption{The best fit output to the data under the background-only hypothesis for the (top left) TL category, (top right) TM category, (bottom left) TT category and (bottom right) all categories, presented together with 68\% \CL uncertainty band for the background model.}
   \label{fig:postfit}
\end{figure*}

The analysis yields no significant excess of events over the SM background prediction. Figure~\ref{fig:postfit} shows the $\mmumu$ distribution in the data of all categories together with the best fit output for the background model, including uncertainties.

\begin{figure}[!hbt]
\centering
	\includegraphics[width=0.48\textwidth]{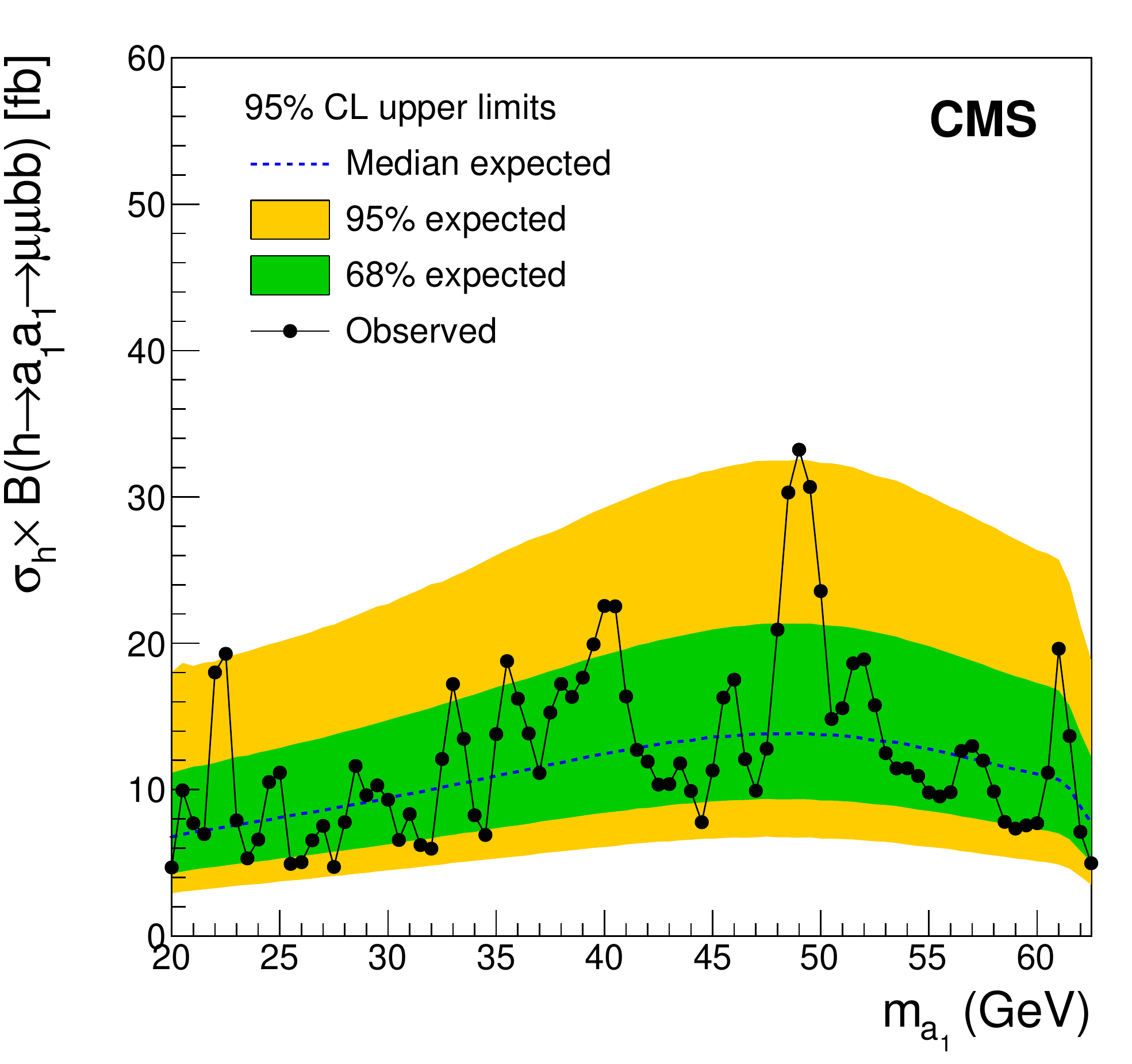}
	\includegraphics[width=0.48\textwidth]{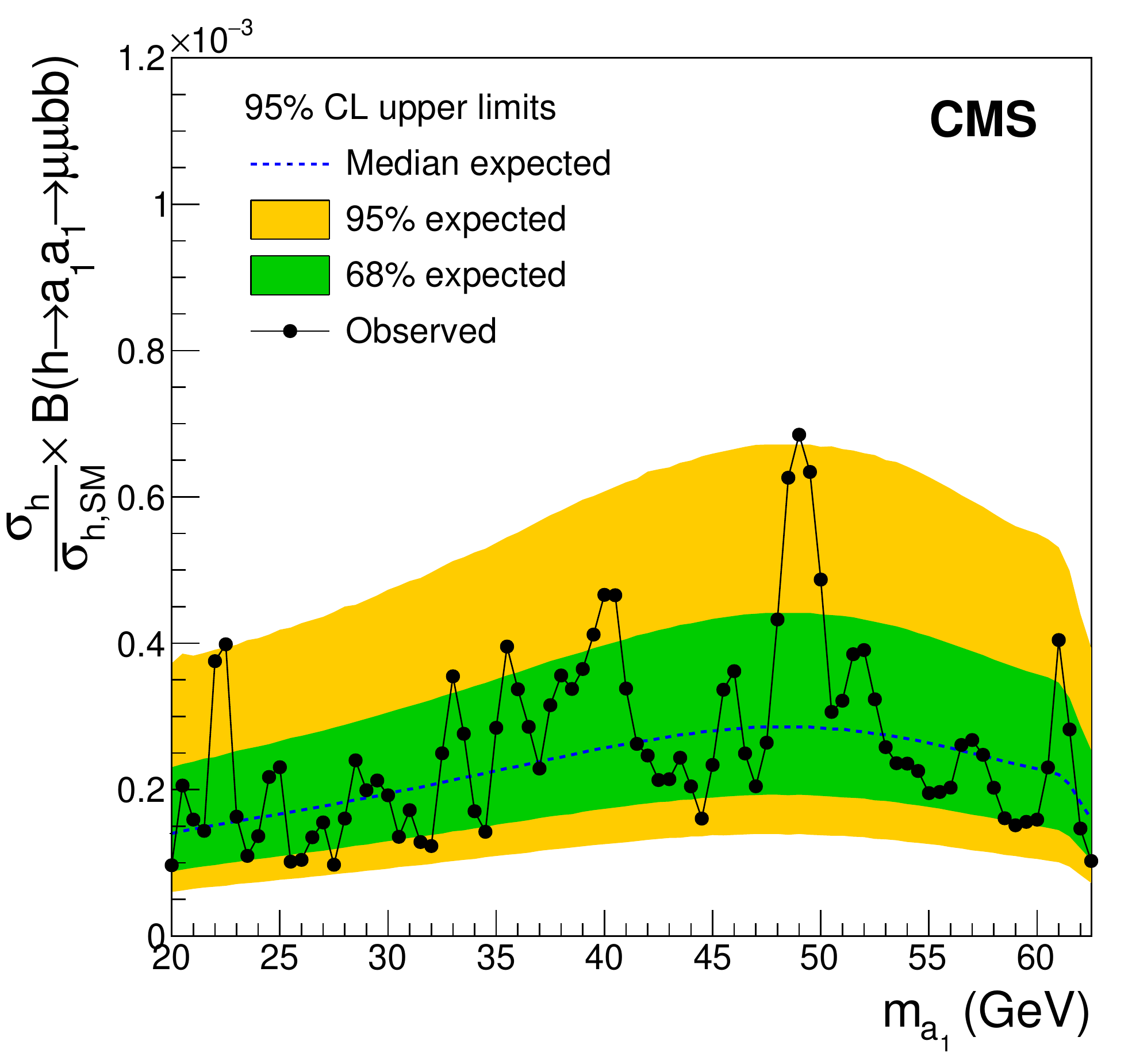}\\
  \caption{Observed and expected upper limits at 95\% \CL on the (\cmsLeft) product of the Higgs boson production cross section and $\mybr$ and (\cmsRight) the branching fraction  as a function of $\ma$. The inner and outer bands indicate the regions containing the distribution of limits located within 68 and 95\% confidence intervals, respectively, of the expectation under the background--only hypothesis.}
  \label{fig:final}
\end{figure}

The upper limit on $\myxsec$ is obtained at 95\% \CL using the CL$_{\mathrm{s}}$ criterion~\cite{Junk1999435,0954-3899-28-10-313} and an asymptotic approximation to the distribution of the profiled likelihood ratio test statistic~\cite{Cowan:2010js}.
Assuming the SM cross sections for the Higgs boson production processes within the theoretical uncertainties, an upper limit is placed on $\mybr$ using the same procedure. Limits are evaluated as a function of $\ma$. The observed and expected limits are illustrated in Fig.~\ref{fig:final} for both cases. Dominant systematic uncertainties are those associated with the \cPqb\ jet identification, followed by the modelling of parton shower and fragmentation. For $\ma=40\GeV$, the \cPqb\ tagging uncertainties arising from LF contamination and JES amount to 17 and 14\%, respectively. The uncertainty arising from the parton shower and fragmentation models is about 7\%. Other uncertainties are below 5\%.

At 95\% \CL, the observed upper limits on $\mybr$ are $(\minBrObsNum$--$\maxBrObsNum)\times 10^{-{4}}$ for the mass range 20 to 62.5\GeV, whilst the expected limits are $(\minBrExpNum$--$\maxBrExpNum)\times 10^{-{4}}$. A similar search from CMS in Run I~\cite{Khachatryan:2017mnf} led to observed upper limits of $(2$--$8)\times 10^{-{4}}$ at 95\% \CL, considering the \cPg\cPg F Higgs boson production and the mass range $25\leq\ma\leq 62.5\GeV$. The corresponding expected limits on the branching fraction at 95\% \CL are $(3$--$4)\times 10^{-{4}}$. At 13\TeV, the \cPg\cPg F Higgs boson production cross section has increased by a factor of about 2.3 over that at 8\TeV, while the production cross section of main backgrounds, Drell--Yan and \ttbar, has increased by a factor of 1.5 and 3.3, respectively. Despite the relative increase in backgrounds, better sensitivity is achieved using improved analysis techniques in Run II.

\begin{figure}[!hbt]
\centering
	\includegraphics[width=0.48\textwidth]{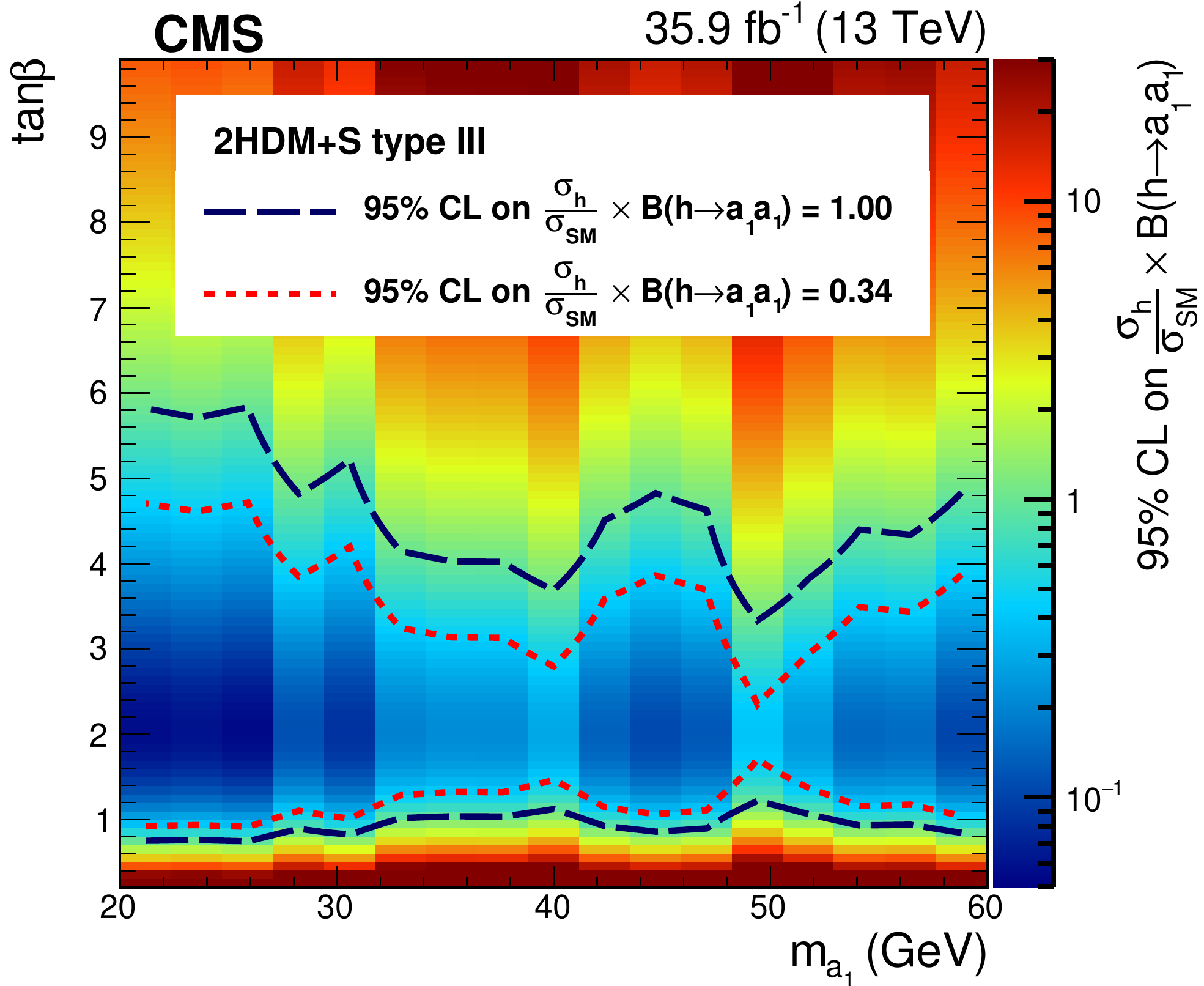}
	\includegraphics[width=0.48\textwidth]{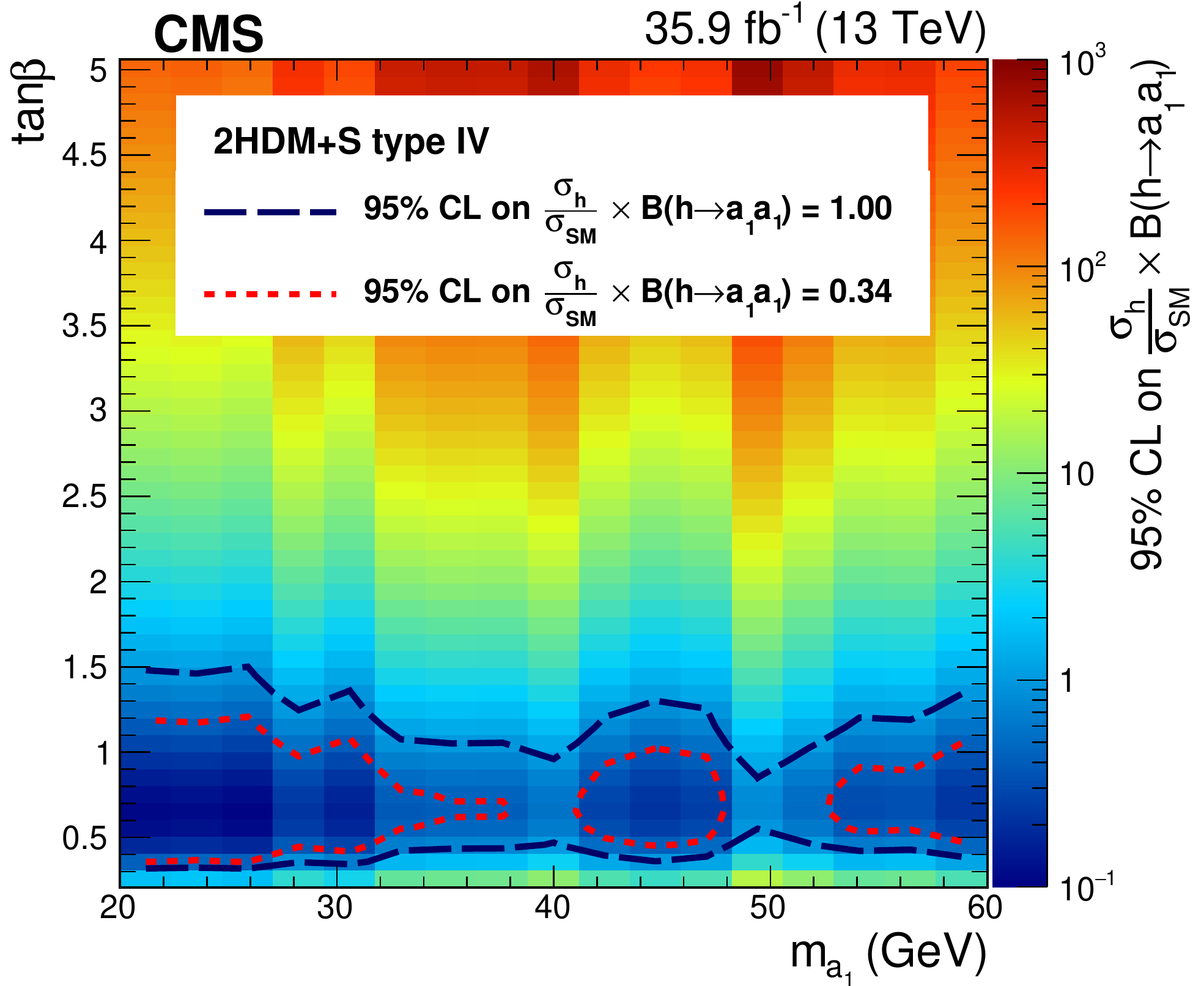}\\
  \caption{Observed upper limits at 95\% \CL on $\mathcal{B}(\Ph\to\ab\ab)$ in the plane of ($\ma,\tanb$) for (\cmsLeft) type-III and (\cmsRight) type-IV 2HDM+S, using only the $\mmbb$ signal.}
  \label{fig:tanbScanmmbb}
\end{figure}

Observed limits on $\mathcal{B}(\Ph\to\ab\ab)$ are shown in Fig.~\ref{fig:tanbScanmmbb} in the plane of ($\ma,\tanb$) for type-III and type-IV 2HDM+S, using only the $\mmbb$ signal. The allowed ranges for $\mathcal{B}(\Ph\to\ab\ab)\leq1$ and $\mathcal{B}(\Ph\to\ab\ab)\leq 0.34$~\cite{ATLASCMSHIGGSCOMBRUNI} are also presented. 

\begin{figure}[!htb]
\centering
	\includegraphics[width=0.48\textwidth]{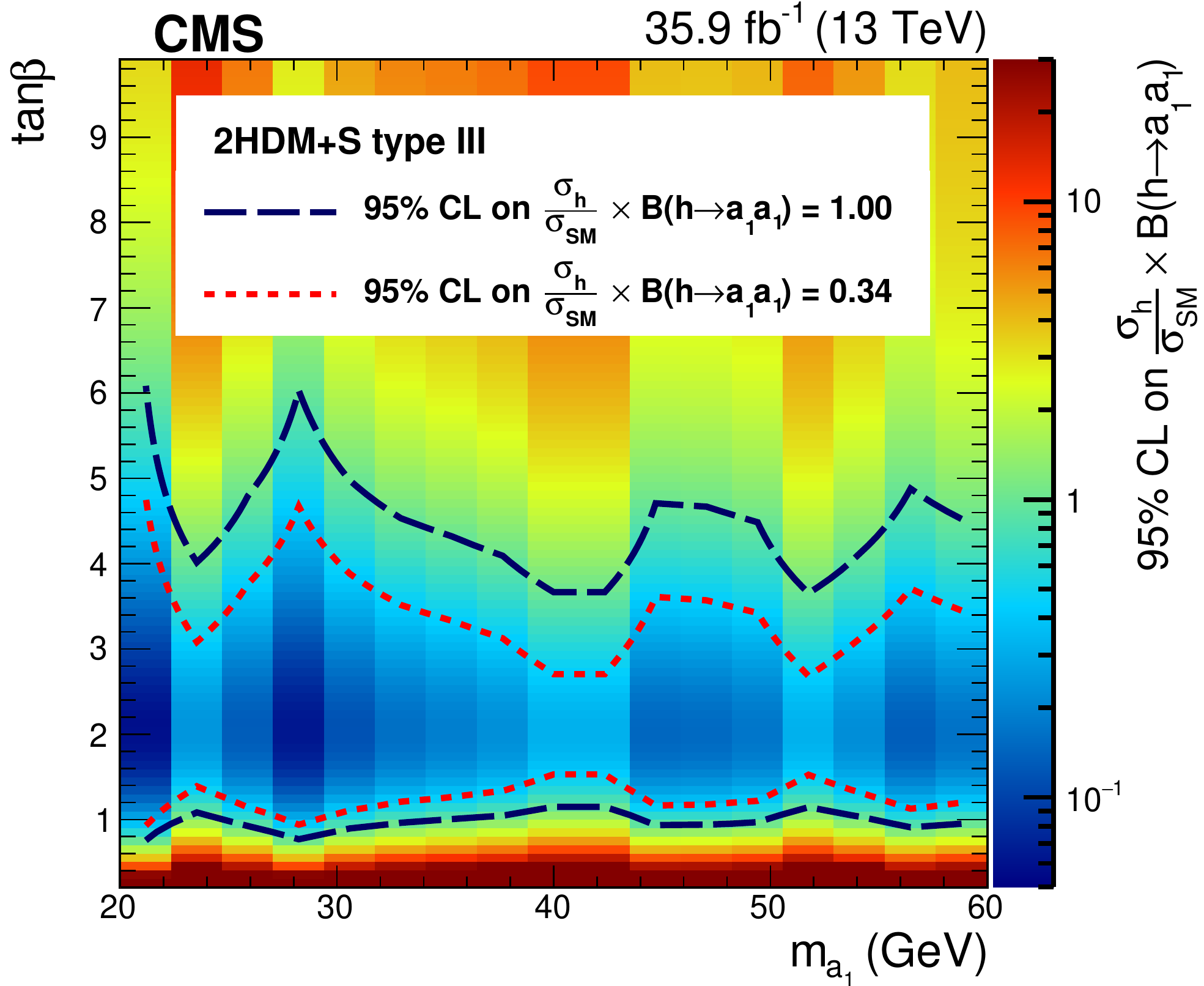}
  \caption{Observed upper limits at 95\% \CL on $\mathcal{B}(\Ph\to\ab\ab)$ in the plane of ($\ma,\tanb$) for type-III 2HDM+S, including $\mmtt$ signal that is misidentified as \mmbb.}
  \label{fig:tanbScanall}
\end{figure}

The effect of including the $\mmtt$ signal is studied in the $(\ma,\tanb)$ plane for the four types of 2HDM+S. For a given $(\ma,\tanb)$ the relevance of $\mmtt$ depends on the ratio $\mathcal{B}(\ab\to\tau\tau)\epsilon^{\text{sel.}}_{\mu\mu\tau\tau}/\mathcal{B}(\ab\to{\cPqb\cPqb})\epsilon^{\text{sel.}}_{\mu\mu{\cPqb\cPqb}}$ as well as the sensitivity of the analysis. Here $\epsilon^{\text{sel.}}$ refers to the acceptance and the selection efficiency of the process. The ratio $\epsilon^{\text{sel.}}_{\mu\mu\tau\tau}/\epsilon^{\text{sel.}}_{\mu\mu{\cPqb\cPqb}}$ is about 1\% in the TL category while it reduces to 0.3 and 0.1\% in the TM and TT categories, respectively. However, because of the increase in the relative branching fraction, the contribution of the $\mmtt$ signal becomes nonnegligible in the type-III 2HDM+S with $\tanb\approx5$. Figure~\ref{fig:tanbScanall} shows the observed limits on $\mathcal{B}(\Ph\to\ab\ab)$ in the ($\ma,\tanb$) plane, including the contribution of $\mmtt$ signal for type-III 2HDM+S. The observed limit contours of $\mathcal{B}({\Ph}\to{\ab\ab})=1.00$ and $\mathcal{B}({\Ph}\to{\ab\ab})=0.34$ are generally extended compared with Fig.~\ref{fig:tanbScanmmbb}\,(\cmsLeft).

\section{Summary}
\label{sec:summary}
A search for the Higgs boson decay to a pair of new pseudoscalars ${\Ph}\to\ab\ab\to\mmbb$, motivated by the next-to-minimal supersymmetric standard model and other extensions to two-Higgs-doublet models, is carried out using a sample of proton-proton collision data corresponding to an integrated luminosity of $\il$ at 13\TeV centre-of-mass energy. No statistically significant excess is found in data with respect to the background prediction. The results of the analysis are presented in the form of upper limits, at 95\% confidence level, on the product of the Higgs boson production cross section and branching fraction, $\myxsec$ as well as on the Higgs boson branching fraction assuming the SM prediction of $\sigma_{\Ph}$. The former ranges between \minTotObs and \maxTotObs\unit{fb}, depending on $\ma$. The corresponding upper limits on the branching fraction are $(\minBrObsNum$--$\maxBrObsNum)\times 10^{-{4}}$ for the mass range of $\lbound\leq\ma\leq \hbound\GeV$. In an analysis performed by ATLAS~\cite{Aaboud:2018esj}, the upper limits on the branching fraction are $(1.2$--$8.4)\times 10^{-4}$. Compared with the similar analysis in Run I~\cite{Khachatryan:2017mnf}, the expected upper limits on the branching fraction are improved by a factor between 1.4 and 1.8 for $25\leq\ma\leq 62.5\GeV$.

\begin{acknowledgments}
We congratulate our colleagues in the CERN accelerator departments for the excellent performance of the LHC and thank the technical and administrative staffs at CERN and at other CMS institutes for their contributions to the success of the CMS effort. In addition, we gratefully acknowledge the computing centres and personnel of the Worldwide LHC Computing Grid for delivering so effectively the computing infrastructure essential to our analyses. Finally, we acknowledge the enduring support for the construction and operation of the LHC and the CMS detector provided by the following funding agencies: BMBWF and FWF (Austria); FNRS and FWO (Belgium); CNPq, CAPES, FAPERJ, FAPERGS, and FAPESP (Brazil); MES (Bulgaria); CERN; CAS, MoST, and NSFC (China); COLCIENCIAS (Colombia); MSES and CSF (Croatia); RPF (Cyprus); SENESCYT (Ecuador); MoER, ERC IUT, and ERDF (Estonia); Academy of Finland, MEC, and HIP (Finland); CEA and CNRS/IN2P3 (France); BMBF, DFG, and HGF (Germany); GSRT (Greece); NKFIA (Hungary); DAE and DST (India); IPM (Iran); SFI (Ireland); INFN (Italy); MSIP and NRF (Republic of Korea); MES (Latvia); LAS (Lithuania); MOE and UM (Malaysia); BUAP, CINVESTAV, CONACYT, LNS, SEP, and UASLP-FAI (Mexico); MOS (Montenegro); MBIE (New Zealand); PAEC (Pakistan); MSHE and NSC (Poland); FCT (Portugal); JINR (Dubna); MON, RosAtom, RAS, RFBR, and NRC KI (Russia); MESTD (Serbia); SEIDI, CPAN, PCTI, and FEDER (Spain); MOSTR (Sri Lanka); Swiss Funding Agencies (Switzerland); MST (Taipei); ThEPCenter, IPST, STAR, and NSTDA (Thailand); TUBITAK and TAEK (Turkey); NASU and SFFR (Ukraine); STFC (United Kingdom); DOE and NSF (USA).

\hyphenation{Rachada-pisek} Individuals have received support from the Marie-Curie programme and the European Research Council and Horizon 2020 Grant, contract No. 675440 (European Union); the Leventis Foundation; the A.P.\ Sloan Foundation; the Alexander von Humboldt Foundation; the Belgian Federal Science Policy Office; the Fonds pour la Formation \`a la Recherche dans l'Industrie et dans l'Agriculture (FRIA-Belgium); the Agentschap voor Innovatie door Wetenschap en Technologie (IWT-Belgium); the F.R.S.-FNRS and FWO (Belgium) under the ``Excellence of Science -- EOS" -- be.h project n.\ 30820817; the Ministry of Education, Youth and Sports (MEYS) of the Czech Republic; the Lend\"ulet (``Momentum") Programme and the J\'anos Bolyai Research Scholarship of the Hungarian Academy of Sciences, the New National Excellence Program \'UNKP, the NKFIA research grants 123842, 123959, 124845, 124850, and 125105 (Hungary); the Council of Science and Industrial Research, India; the HOMING PLUS programme of the Foundation for Polish Science, cofinanced from European Union, Regional Development Fund, the Mobility Plus programme of the Ministry of Science and Higher Education, the National Science Center (Poland), contracts Harmonia 2014/14/M/ST2/00428, Opus 2014/13/B/ST2/02543, 2014/15/B/ST2/03998, and 2015/19/B/ST2/02861, Sonata-bis 2012/07/E/ST2/01406; the National Priorities Research Program by Qatar National Research Fund; the Programa Estatal de Fomento de la Investigaci{\'o}n Cient{\'i}fica y T{\'e}cnica de Excelencia Mar\'{\i}a de Maeztu, grant MDM-2015-0509 and the Programa Severo Ochoa del Principado de Asturias; the Thalis and Aristeia programmes cofinanced by EU-ESF and the Greek NSRF; the Rachadapisek Sompot Fund for Postdoctoral Fellowship, Chulalongkorn University and the Chulalongkorn Academic into Its 2nd Century Project Advancement Project (Thailand); the Welch Foundation, contract C-1845; and the Weston Havens Foundation (USA). \end{acknowledgments}

\bibliography{auto_generated}

\cleardoublepage \appendix\section{The CMS Collaboration \label{app:collab}}\begin{sloppypar}\hyphenpenalty=5000\widowpenalty=500\clubpenalty=5000\input{HIG-18-011-authorlist.tex}\end{sloppypar}
\end{document}

%% file: HIG-18-011-authorlist.tex
\vskip\cmsinstskip
\textbf{Yerevan Physics Institute, Yerevan, Armenia}\\*[0pt]
A.M.~Sirunyan, A.~Tumasyan
\vskip\cmsinstskip
\textbf{Institut f\"{u}r Hochenergiephysik, Wien, Austria}\\*[0pt]
W.~Adam, F.~Ambrogi, E.~Asilar, T.~Bergauer, J.~Brandstetter, M.~Dragicevic, J.~Er\"{o}, A.~Escalante~Del~Valle, M.~Flechl, R.~Fr\"{u}hwirth\cmsAuthorMark{1}, V.M.~Ghete, J.~Hrubec, M.~Jeitler\cmsAuthorMark{1}, N.~Krammer, I.~Kr\"{a}tschmer, D.~Liko, T.~Madlener, I.~Mikulec, N.~Rad, H.~Rohringer, J.~Schieck\cmsAuthorMark{1}, R.~Sch\"{o}fbeck, M.~Spanring, D.~Spitzbart, W.~Waltenberger, J.~Wittmann, C.-E.~Wulz\cmsAuthorMark{1}, M.~Zarucki
% ...
\vskip\cmsinstskip
% ...
1:  Also at Vienna University of Technology, Vienna, Austria\\
% ...
